\documentclass[lettersize,journal]{IEEEtran}
\usepackage{amsmath,amsfonts}
\usepackage{algorithmic}
\usepackage{algorithm}
\usepackage{array}
\usepackage{textcomp}
\usepackage{stfloats}
\usepackage{url}
\usepackage{verbatim}
\usepackage{graphicx}
\usepackage{cite}
\usepackage{subcaption}
\usepackage{xcolor}
\usepackage{float}
\usepackage[bottom]{footmisc} 

\usepackage{xcolor,soul}
\usepackage{xspace}
\usepackage{enumitem}
\usepackage{titlecaps}
\usepackage{tabularx}
\usepackage{balance}
\usepackage{color}
\usepackage{comment}
\usepackage{grffile} 
\newcommand{\eg}{{\em e.g.,}\ }
\newcommand{\ie}{{\em i.e.,}\ }
\newcommand{\sol}{{\em IoT-AD}\xspace}
\newcommand{\solnospace}{{\em IoT-AD}}

\newcommand{\revone}{\color{black}}

\begin{document}

\makeatletter
\def\ps@IEEEtitlepagestyle{%
  \def\@oddfoot{\mycopyrightnotice}%
  \def\@evenfoot{}%
}
\def\mycopyrightnotice{%
  {\footnotesize \textcolor{red}{\begin{tabular}[t]{@{}l@{}} This paper has been accepted for publication by the IEEE Internet of Things Journal. © 2023 IEEE. Personal use of this material is permitted. Permission \\ from IEEE must be obtained for all other uses, in any current or future media, including reprinting/republishing this material for advertising or promotional \\ purposes, creating new collective works, for resale or redistribution to servers or lists, or reuse of any copyrighted component of this work in other works.\end{tabular}}}
  \gdef\mycopyrightnotice{}
}

\title{IoT-AD: A Framework To Detect Anomalies Among Interconnected IoT Devices}
%

\author{Hasniuj Zahan, Md Washik Al Azad, Ihsan Ali, and Spyridon Mastorakis
\IEEEcompsocitemizethanks{
    \IEEEcompsocthanksitem{Hasniuj Zahan (email: hzahan@unomaha.edu) and Ihsan Ali (email: ihsanali@ieee.org) are with the Department of Computer Science, University of Nebraska at Omaha, Nebraska, USA. Md Washik Al Azad (email: malazad@nd.edu) and Spyridon Mastorakis (corresponding author, email: mastorakis@nd.edu) are with the Department of Computer Science and Engineering, University of Notre Dame, Indiana, USA. } 
    
}
}



\maketitle

\begin{abstract}

In an Internet of Things (IoT) environment (\eg smart home), several IoT devices may be available that are interconnected with each other. In such interconnected environments, a faulty or compromised IoT device could impact the operation of other IoT devices. In other words, anomalous behavior exhibited by an IoT device could propagate to other devices in an IoT environment. In this paper, we argue that mitigating the propagation of the anomalous behavior exhibited by a device to other devices is equally important to detecting this behavior in the first place. In line with this observation, we present a framework, called IoT Anomaly Detector (\sol), that can not only detect the anomalous behavior of IoT devices, but also limit and recover from anomalous behavior that might have affected other devices. We implemented a prototype of \sol, which we evaluated based on open-source IoT device datasets as well as through real-world deployment on a small-scale IoT testbed we have built. \textcolor{black}{We have further evaluated \sol in comparison to prior relevant approaches.} Our evaluation results show that \sol can identify anomalous behavior of IoT devices in less than 2.12 milliseconds and with up to 98\% of accuracy.

\end{abstract}



\section{Introduction}

With the proliferation of Internet of Things (IoT), different IoT environments, such as smart homes, have become a reality. Such environments may consist of a number of IoT devices from different vendors. These devices may be interconnected to each other and adhere to conditions defined by users (\eg users select a temperature of their preference for their smart thermostat). In this context, there are several IoT device controller platforms, such as IFTTT, Samsung SmartThings, and the Apple Homekit\footnote{{{IFTTT: \url{https://www.ifttt.com/}, Samsung SmartThings: \url{https://www.smartthings.com/}, Apple Homekit: \url{https://developer.apple.com/apple-home/}.}}}, which integrate the services offered by IoT devices of different vendors. However, when these devices exhibit behavior that does not follow the conditions defined by users or report inaccurate readings, \textcolor{black}{this behavior is defined as an anomaly}. Given the interconnected nature of IoT devices, an anomaly occurred/triggered by an IoT device can propagate and affect other IoT devices. 

\begin{figure}
\centering
\includegraphics[width=1\linewidth]{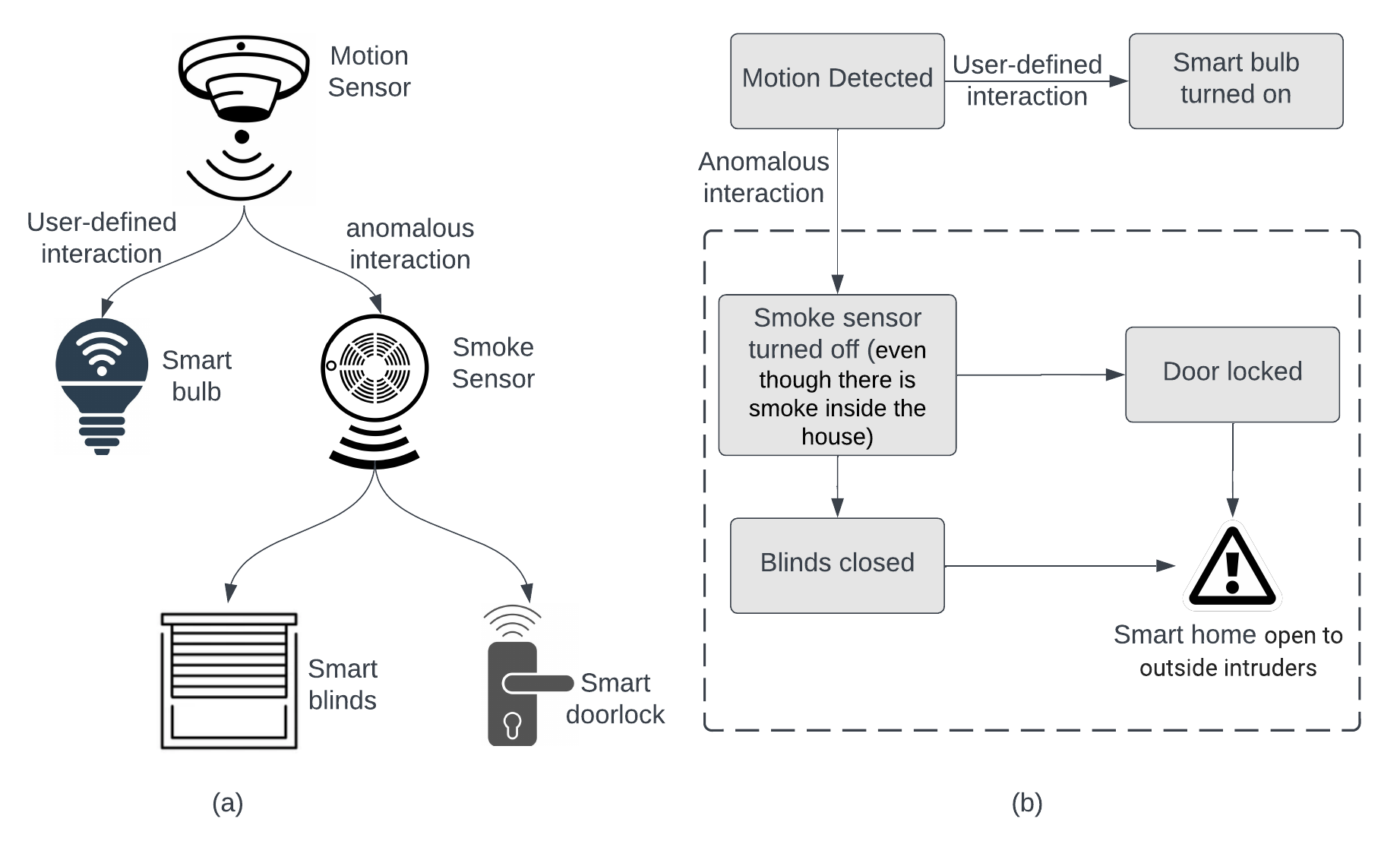}
\caption{(a) Interactions among IoT devices; (b) Propagation of an anomaly across IoT devices.}
\label{fig:case3}
\end{figure}

We present an anomaly propagation scenario in a smart home (Figure \ref{fig:case3}). In this scenario, a faulty or compromised motion sensor interacts with and turns on a smoke sensor even if no smoke is actually detected in the house. As a result, the smoke detector interacts with sensors that open the windows and unlock the door of the house, so that residents can escape the smoke and exit the house safely. In this context, the anomalous behavior of a faulty or compromised motion sensor has propagated and affected a smoke sensor, smart window sensors, \textcolor{black}{and a} smart lock, resulting in a situation that the residents' house is open to outside intruders.

Prior research has identified different types of anomalies and \textcolor{black}{has proposed methods to identify these anomalies}{\cite{DBLP:journals/corr/abs-1910-03750, 8416520, Bakar2016ActivityAA, cauteruccio2021framework}}. The majority of prior research has considered: (i) packet-level anomalies related to the network traffic generated by IoT devices (\eg while communicating with controller platforms, cloud services, or other IoT devices); and (ii) state/behavior anomalies based on the analysis of the IoT device behavior over time. However, prior research has focused on anomaly detection methods, but did not investigate mechanisms to: (i) mitigate the effects of anomalies that propagate from one IoT device to another, thus affecting the state/behavior of multiple devices in an IoT environment; and (ii) enabling affected IoT devices to recover from such propagated anomalies.

\noindent\textbf{\textcolor{black}{Motivation and research questions:}}
\textcolor{black}{{\revone Solely detecting IoT device anomalies and isolating devices that experienced anomalies} may not be sufficient in real-world scenarios, such as smart homes or buildings. {\revone Given the interconnected nature of real-world IoT environments, anomalies can propagate among IoT devices before they can be identified}. In other words, an anomaly caused by an IoT device can affect other IoT devices. To this end, mechanisms are necessary to mitigate the effects of anomalies that \textcolor{black}{can propagate among IoT devices}. To address this challenge, in this paper, we investigate the following research questions:}

\begin{itemize}

\item How can we realize a framework \textcolor{black}{that detects} packet-level anomalies and anomalies related to the interactions among IoT devices?


\item How can such a framework enable \textcolor{black}{affected} IoT devices to recover from propagated anomalies and revert to their last known stable state as if the detected anomalies had never {\revone occurred}?

\end{itemize}


\noindent\textbf{Our contributions:} In this paper, we present IoT Anomaly Detector (\sol), a framework that: (i) detects packet-level anomalies \textcolor{black}{and interaction} anomalies, which can propagate across IoT devices; and (ii) mitigates the effects of propagated anomalies, enabling IoT devices to recover from propagated anomalies. \sol essentially acts as an extended IoT device controller framework, which is able to detect anomalies in IoT environments and help associated IoT devices recover from these anomalies. The contribution of our work is two-fold:  


\begin{itemize}

\item We present the \sol design, which features mechanisms to detect packet-level anomalies and anomalies related to the interactions among IoT devices, maintain the state of IoT devices over time, and ultimately enable IoT devices to recover from anomalies that may \textcolor{black}{have propagated among them.}


\item We implement an \sol prototype, which we evaluate based on open-source IoT device datasets as well as through real-world deployment on a small-scale IoT testbed we have built. \textcolor{black}{We further evaluate this prototype in comparison to prior relevant approaches for anomaly detection in IoT environments.} Our evaluation results demonstrate that \sol is a light-weight framework that can identify IoT device anomalies in less than 2.12 milliseconds (ms) and with up to 98\% of accuracy.

 
\end{itemize}

The rest of our paper is organized as follows. In Section \ref{subsec:background}, we discuss a brief background on IoT anomaly detection and mitigation, and present prior related work. In Section \ref{sec:overview}, we present our threat model and a design overview of \sol. In Section \ref{sec:design}, we present the design of \sol, in Section \ref{sec:eval}, we present the evaluation of \sol, and, in Section~\ref{sec:discussion}, we discuss extensions and limitations of the \sol design. Finally, in Section \ref{sec:concl}, we conclude our paper.





\section{Background and Prior Related Work}
\label{subsec:background}

In this section, we present a brief background on IoT anomaly {  detection} and mitigation, and discuss related work.

\subsection{Anomaly Detection of IoT Devices}

The purpose of anomaly detection is to identify patterns whose behavior is considered atypical compared to typical ones. \textcolor{black}{In prior research,} different approaches have been used to detect anomalies in an IoT environment, such as traffic analysis, packet-level signatures, and semantics-based models. Homonit~\cite{10.1145/3243734.3243820} \textcolor{black}{is a framework, mainly for the Samsung SmartThings platform,} which can detect anomalies by monitoring encrypted traffic patterns of smart devices’ activities, and comparing these patterns with expected behaviors inferred from the source code. \textcolor{black}{HADES-IoT \cite{breitenbacher2021hades} is a light-weight host-based IoT anomaly detection framework that can proactively detect and prevent the execution of unauthorized functions on IoT devices.} {  IoT-Praetor}~\cite{9143140} leverages a device usage description model to profile different communication and interaction behaviors among IoT devices to automatically detect anomalies in real time. A semantics-based approach, called HAWatcher~\cite{263888}, was proposed by Wang \textit{\textcolor{black}{\textit{et al.}}}, where semantic information of IoT devices, such as applications, device types, installation locations, and event logs, is used to generate correlations. Subsequently, a shadow execution simulation is executed based on these correlations to simulate \textcolor{black}{a smart home environment in parallel with the actual smart home environment. The behavior of the real and} the simulated environment is compared to detect behavioral anomalies.

HomeSnitch~\cite{10.1145/3317549.3323409} uses the perspective of software-defined networking to track communication between devices and servers, classify device actions, and find anomalous behavior. PingPong~\cite{https://doi.org/10.48550/arxiv.1907.11797} is a tool that extracts packet-level signatures for device events (\eg a smart plug turning ON/OFF) from network traffic to identify anomalies in smart home networks. 
Orpheus~\cite{10.1145/3134600.3134640} is an anomaly detection tool that uses event logs and system logs to detect data-oriented exploits and different runtime attacks. Xu \textcolor{black}{\textit{et al.}}~\cite{10.1002/sec.1569} proposed a system that analyzes home network traffic using a bloom filter to detect anomalous traffic behavior. \textcolor{black}{Li \textit{\textcolor{black}{\textit{et al.}}} proposed a light-weight statistical learning approach for IoT devices where different system information, such as CPU and memory usage, and network throughput, can be used to detect anomalies~\cite{li2019system}. The same authors also proposed a technique to detect attacks against IoT devices based on the energy consumption of different system components (\eg CPU, network) \cite{li2019enhanced}.} These systems mostly used either event logs to infer device activity or packet-level signatures to find out deviations from the usual traffic patterns. 
 
{\revone Furthermore, anomaly detection frameworks have been proposed based on features of the physical layer, such as the Received Signal Strength Indicator (RSSI), the Power Spectral Density (PSD), and the Signal-to-Noise Ratio (SNR). Martins \textit{et al.} used raw RSSI data to extract the silence and activity periods of IoT devices and detect anomalies based on this information~\cite{martins2020physical}. Tang \textit{et al.} proposed an anomaly detection framework for wireless sensor networks, which is also based on RSSI data~\cite{tang2007rssi}. Rajendran \textit{et al.} proposed SAIFE, which analyzes the PSD of the wireless spectrum within an IoT enviroment to detect anomalies~\cite{rajendran2018saife}. The challenges of detecting anomalies based on physical layer information are the following: (i) the collected signals may be susceptible to noise; and (ii) the features used to detect anomalies may depend on the distance between the sender and the receiver.}


\subsection{Machine Learning for IoT Device Anomaly Detection}

Due to recent technological advancements, machine learning has become a powerful technique for the detection of anomalies by identifying irregular data patterns. Yin \textcolor{black}{\textit{et al.}}~\cite{4358934} and Jakkula \textcolor{black}{\textit{et al.}}~\cite{Jakkula2011DetectingAS} used {  a one-class support vector machine} to identify activity patterns in a smart home using a sensor-based dataset. Ramapatruni \textcolor{black}{\textit{et al.}}~\cite{8819458} used a hidden Markov model to train network-level data of a smart home in order to detect anomalies. Fahad \textcolor{black}{\textit{et al.}}~\cite{Fahad2015AnomaliesDI} used the density based spatial clustering algorithm for user activity recognition, while Trimananda \textcolor{black}{\textit{et al.}}~\cite{https://doi.org/10.48550/arxiv.1907.11797} used the same algorithm to cluster packet pairs \textcolor{black}{and detect anomalies in network traffic.} With the combination of two different techniques, the principal component analysis and a sliding window algorithm, Wu \textcolor{black}{\textit{et al.}}~\cite{Wu2019TWSeeHA} can identify user activities by analyzing WiFi signals. Branch \textcolor{black}{\textit{et al.}}~\cite{Branch_Giannella_Szymanski_Wolff_Kargupta_2013} used a K-NN machine learning method for outlier detection, while Narudin {  \textit{et al.}}~\cite{10.1007/s00500-014-1511-6} used both naive bayes and random forest machine learning algorithms to detect anomaly-based malware. 

\subsection{Anomaly Mitigation in Smart Homes}

To reduce the effects caused by an anomaly, it is necessary to create prevention and recovery methods. 
Noah \textcolor{black}{\textit{et al.}}~\cite{journals/corr/abs-1812-00955} introduced a stochastic traffic padding algorithm that shapes the traffic \textcolor{black}{to hide the metadata and} keep user activities private. Gaurav \textcolor{black}{\textit{et al.}}~ \cite{10.1145/3510547.3517920} proposed a framework to enhance the security of a smart home that uses attribute based access control. The advantage of using attribute based access control is the ability to specify fine-grained security policies and consider environmental conditions to make access decisions. Yamauchi \textcolor{black}{\textit{et al.}}~\cite{9040414} proposed a system that detects anomalous events by analyzing event sequences and dropping packets associated with anomalous events. 



\begin{table*}[t]
\centering
\textcolor{black}{
\caption{Comparison of design properties of \sol and other prior related work (Y: Yes, N: No, L: Limited).}
\resizebox{0.95\textwidth}{!}{%
\label{Tab:related_work_comps}
\begin{tabular}{|c|c|c|c|c|c|c|}
\hline
\textbf{}                                                                                  & \multicolumn{1}{l|}{Homonit{  ~\cite{10.1145/3243734.3243820}}} & IoT-Praetor{  ~\cite{9143140}} & HAWatcher{  ~\cite{263888}} & PingPong{  ~\cite{https://doi.org/10.48550/arxiv.1907.11797}} & \begin{tabular}[c]{@{}c@{}} Yamauchi {  \textit{et al.}}~\cite{9040414} \end{tabular}  & \multicolumn{1}{l|}{\textit{\textbf{\sol}}} \\ \hline
\begin{tabular}[c]{@{}c@{}}Traffic anomaly \\ detection \end{tabular}                                                                      & Y       & Y       & N          & L  & N                                                  & \textit{\textbf{Y}}                              \\ \hline
\begin{tabular}[c]{@{}c@{}}Interaction anomaly \\ detection \end{tabular}                                                                       & L       & Y       & Y          & N   & Y                                                 & \textit{\textbf{Y}}                              \\ \hline
\begin{tabular}[c]{@{}c@{}} Stable state \\ recovery \end{tabular}                                                                               & N       & N       & N          & N  & L                                                   & \textit{\textbf{Y}}                              \\ \hline
\end{tabular}
}
}
\vspace{-0.3cm}
\end{table*}

\noindent\textcolor{black}{{\textbf{How is \sol different from prior related work:}  \textcolor{black}{In Table \ref{Tab:related_work_comps}, we present a comparison of the \sol's design properties to prior work. The majority of prior work focused on:} (i) { anomalous behavior detection of IoT devices} using traffic analysis; and/or (ii) analyzing the interaction patterns among the interconnected IoT devices \textcolor{black}{in an IoT environment}. For example, PingPong\textcolor{black}{\cite{https://doi.org/10.48550/arxiv.1907.11797}} provides a way to \textcolor{black}{generate} signatures of IoT events by analyzing the packet sequences, which can be used to identify anomalous events, while HAWatcher\textcolor{black}{\cite{263888}} generates hypothetical correlations among events of different IoT devices (\ie interactions among devices) to detect anomalies. IoT-Praetor\textcolor{black}{\cite{9143140}} and Homonit\textcolor{black}{\cite{10.1145/3243734.3243820}} are specific to the Samsung SmartThings platform. A drawback of prior work is that these frameworks did not provide any mechanisms to help IoT devices recover from anomalies and revert to their last known stable states. {\revone To this end, \sol not only utilizes traffic and interaction analysis to detect anomalies that can propagate among interconnected IoT devices but also provides mechanisms that enable affected IoT devices to recover from propagated anomalies and revert to their last known stable state as if the detected anomalies had not happened}.}}


\section{Threat Model and Design Overview}

In this section, we first describe the threat model that we consider in the context of \sol and we then present an overview of the \sol design.

\label{sec:overview}
\subsection{Threat Model}

As IoT devices are resource-constrained and manufactured for minimal and specific functions to reduce cost and complexity, these devices suffer from malfunctions and security vulnerabilities~\cite{8753563, Andrea2015InternetOT, DBLP:journals/corr/abs-1808-02125, 9099839}. As a result, IoT devices may introduce anomalies into an IoT environment. In this paper, we have considered anomalies that are caused by: (i) device malfunctions; and (ii) devices compromised due to malicious attacks. We further assume that an IoT device controller platform that may be available will operate legitimately.

\noindent\textbf{Anomalies due to IoT device malfunction:} As IoT devices have limited power and computing resources, \textcolor{black}{they are prone to malfunction. In most cases, there are no built-in malfunction detection mechanisms.} As a result, it is challenging to identify a faulty or malfunctioning device in a timely manner and repair or replace the device. Device malfunctions can be either software-based or hardware-based.


Ghost commands, command failures, delays in status updates, and event losses occur as a result of software-based device malfunctions. Such malfunctions can happen due to a system crash, network connectivity issues, or bugs in application or operating system code. Hardware-based malfunctions can create false events and command failures. For example, a defective motion sensor can erroneously detect human presence when there are no humans around. Another example of hardware-based malfunctions is command failures due to \textcolor{black}{melted capacitors and faulty circuit boards. Both types of malfunctions (software-based and hardware-based)} can lead to unexpected communication patterns (\eg sending updates irregularly instead of periodically) or unexpected interactions with other IoT devices (\eg a thermostat turns on the fan even though the temperature sensor indicates that the room temperature has not changed). Both types of unexpected patterns can be used to identify anomalies generated by an IoT device.

\noindent\textbf{Anomalies due to compromised IoT devices:} Another reason for IoT device anomalies may be the fact that an IoT device has been compromised. Given that IoT devices are resource-constrained, malicious actors can compromise and gain access to these devices. For example, attackers can exploit weak usernames and passwords chosen by users, network vulnerabilities, or create malicious applications that users can install on their IoT devices. If attackers gain access to an IoT device, they can compromise the behavior of the device (\eg produce bogus readings), so that the device interacts with other devices and the anomaly propagates. In addition, attackers can use a compromised IoT device \textcolor{black}{to access other IoT devices}, which happened during the Mirai botnet attack{ ~\cite{antonakakis2017understanding}}. Since the behavior/patterns of a compromised device will typically deviate from its regular operation patterns, such changes of device operation patterns can be used to identify anomalies.

\subsection{\sol Design Overview}

\begin{figure}
\centering
\includegraphics[width=1\linewidth]{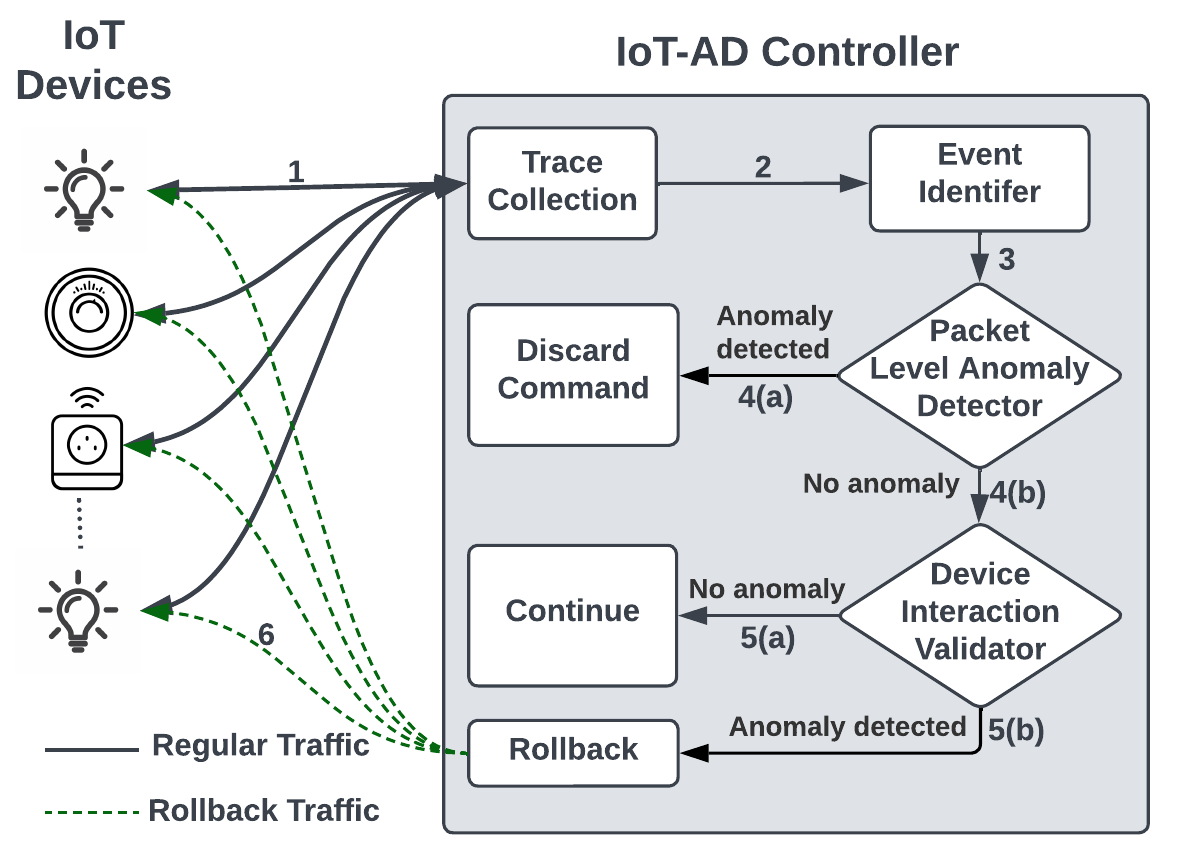}
\caption{\textcolor{black}{Operation workflow of \sol.}}
\label{fig:System}
\end{figure}

In the context of \sol, we consider an IoT environment (\eg a smart home), where IoT devices are interconnected through a controller. The controller realizes the intelligence of the IoT environment, communicates with IoT devices, analyzes network traffic and interactions among IoT devices, and instructs devices to change their current state as necessary. \sol operates in five steps (Figure~\ref{fig:System}). \textcolor{black}{The controller is responsible for executing these steps, which we present below:}

\noindent\textbf{Device monitoring:} The controller monitors all IoT devices and receives measurements/readings from them over time. These measurements/readings are transmitted to the controller over the local network through a series of packet exchanges.

\noindent\textbf{Event identifier and device status update:} Each IoT device communicates with the controller through a series of network packet exchanges (application layer protocols). Given the heterogeneous nature of IoT devices (\eg different vendors, different software stacks), each device may implement different application layer protocols to communicate with the controller. For each device, the controller identifies events (as communicated {by devices}) that affect the status of each device and maintains a log where it records the status of each device over time. For instance, when a smart light bulb turns on (this is the event in our example), the bulb will communicate with the controller and the controller will update the status of this device to ``on''.



\noindent\textbf{Packet-level anomaly detection:}
As IoT devices communicate through the controller, the controller monitors the generated network traffic and produces packet-level signatures for different events. These signatures are used by the controller over time to identify packet-level (network traffic) anomalies. If a packet-level anomaly is detected, the controller will discard the event.

\noindent\textbf{Device interaction validation:} IoT devices interact with each other via the controller. We use the term ``device interaction'' to refer to events triggered by a measurement/reading of an IoT device that affects one or more other devices. For example, if a motion sensor detects human presence in a room, based on the time of the day, \textcolor{black}{it may trigger a smart light to turn on.} 
The controller will validate the interactions among devices and identify anomalies related to these interactions. Such an anomaly may occur, for example, when a motion sensor attempts to trigger a smoke detector when it detects motion in a room of the house, since this interaction is not meaningful.


\noindent\textbf{Rollback:} Once an anomaly is detected, the controller identifies how much this anomaly has ``propagated'' among IoT devices. For example, a motion detector may attempt to trigger a smoke detector (meaningless interaction), which can subsequently unlock the smart lock of a house's front door and windows (meaningful action, since residents will need to exit the house in the case of a fire). In other words, the controller will identify which devices have been affected by an interaction anomaly and revert the state of the affected devices back to the most recent stable state.

\section{\sol Design}
\label{sec:design}

\textcolor{black}{In this section, we present the components of the \sol design in detail. A list of symbols and abbreviations used in the design of \sol is provided in Table~\ref{tab:symbols}.}

\begin{table}[b]
\centering
\textcolor{black}{
\caption{\textcolor{black}{List of symbols and abbreviations used in the design of \sol.}}
\label{tab:symbols}
\begin{tabular}{|l|l|}
\hline
\textbf{Symbol/Abbreviation} & \textbf{Description}               \\ \hline
L2 headers   & Link layer header fields \\ \hline
L4 headers   & Transport layer header fields \\ \hline
L3 headers   & Network layer header fields \\ \hline
SYN          & TCP synchronization flag   \\ \hline
ACK          & TCP acknowledgement flag   \\ \hline
FIN          & TCP finish flag            \\ \hline
TCP cmd      & TCP command            \\ \hline
TCP rsp      & TCP response           \\ \hline
Event $e$          & An event detected by the controller           \\ \hline
\end{tabular}}
\end{table}

\subsection{Device Monitoring}

In \sol, the controller is the entity that conducts a number of operations (\eg packet-level anomaly detection, device interaction validation). To this end, existing IoT devices with adequate resources (\eg smart TVs, smart refrigerators), resourceful WiFi routers, or smart hubs\footnote{{Examples of a smart TV, a refrigerator, a WiFi router, and a smart hub:} \begin{itemize}[noitemsep]
\item \textcolor{black}{Smart TV: \url{https://developer.samsung.com/smarttv/develop/specifications/general-specifications.html}
\item \textcolor{black}{Smart refrigerator: \url{https://www.samsung.com/us/home-appliances/refrigerators/all-refrigerators/}}
\item \textcolor{black}{WiFi router: \url{https://www.asus.com/us/Networking-IoT-Servers/WiFi-Routers/ASUS-Gaming-Routers/RT-AX88U/}} 
\item \textcolor{black}Smart hub: \url{https://store.google.com/us/product/nest_wifi_specs}}
\end{itemize}}
can act as controllers. All devices in an IoT environment are connected via wireless to a controller, which is responsible for monitoring all devices within this environment and for collecting measurements/readings from these devices over time. Depending on the nature of an IoT device, the state information of devices can be updated to the controller either periodically or whenever a state change occurs. For instance, the temperature readings of a thermostat can be updated to the controller after a certain time interval. Another example may be that when an event occurs, \textcolor{black}{such as a smart light bulb turning on or off,} the state of the light bulb will be updated to the controller.

Apart from monitoring the states of IoT devices, the controller can also analyze packets that are exchanged through the controller among the devices in chronological order.  When a user, for example, tries to turn on a smart device (\eg a smart bulb) through an application, multiple packets are exchanged between the smart device and the application. The controller monitors various L2-L4 packet headers (\eg IP addresses, TCP/UDP ports, packet lengths) and records the timestamp of each packet. The headers of exchanged packets are used to identify the communication patterns among devices and keep an event log for each device. 



\subsection{Event Identifier and Device Status Updates} 

An event is a series of packet exchanges between two or more IoT devices within the local network that have interdependencies (\eg a motion sensor can turn on a light bulb \textcolor{black}{or even multiple light bulbs and plugs)}. When a device communicates with a cloud server for a particular service, such as searching for a firmware update, sending device information, receiving control commands from a user application, \textcolor{black}{this communication can be also identified as an event.} 

\begin{figure}
\centering
\includegraphics[width=1\linewidth]{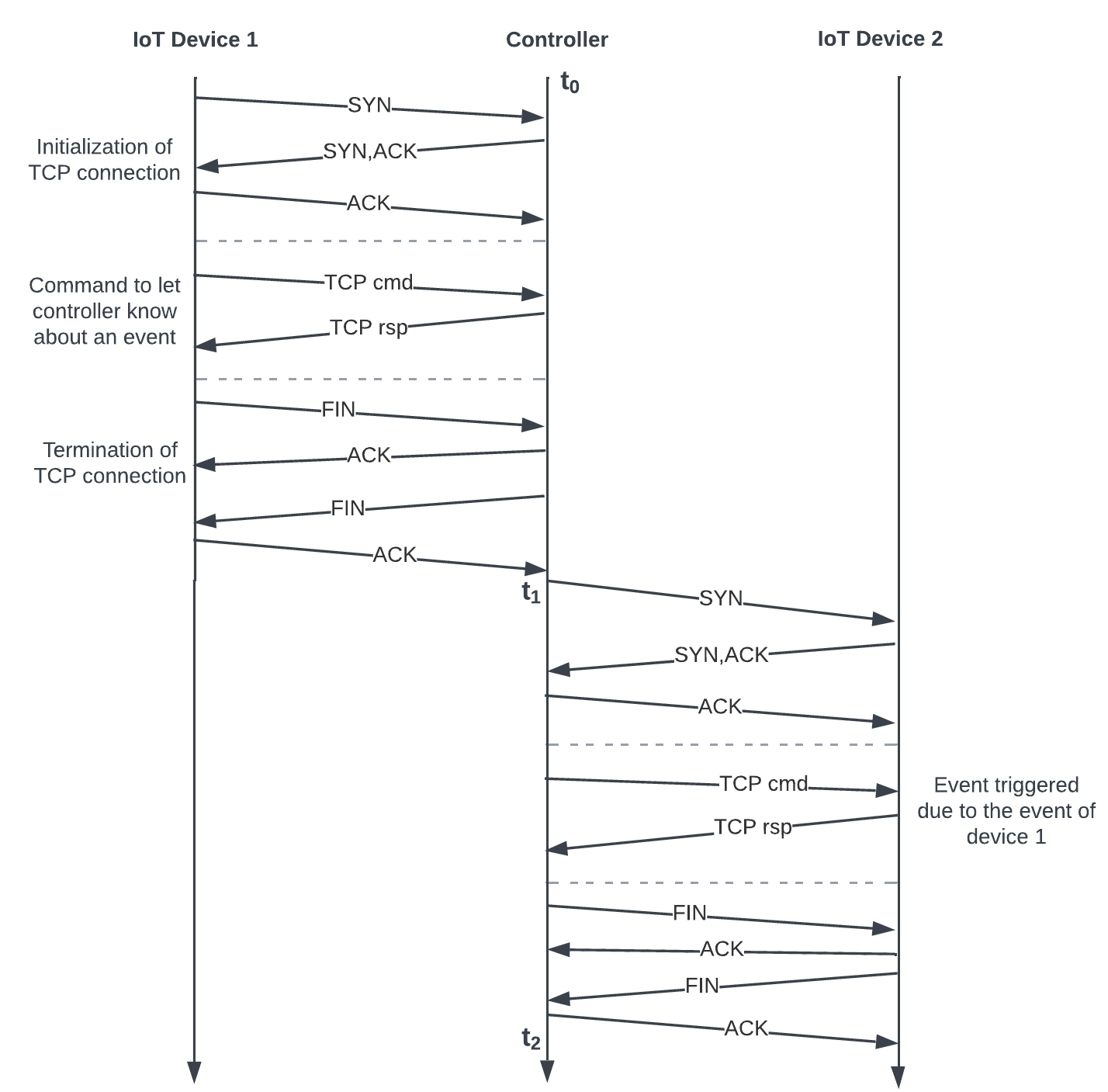}
\caption{An example of the network traffic \textcolor{black}{generated for communication} between two IoT devices through the \sol controller.}
\label{fig:flowdiagram}
\vspace{-0.3cm}
\end{figure}

Figure~\ref{fig:flowdiagram} illustrates an example of packet exchanges between two IoT devices through the controller in \sol. First, a TCP connection is established between device 1 and the controller. The device then sends a command related to an event (\eg a motion sensor detecting motion in a room) to the controller. After the command is successfully received by the controller, the TCP connection with device 1 is closed, and the controller follows the same process to communicate a command with device 2. This command is triggered by the event of device 1 (\eg a motion sensor that detected motion in a room, which in turn activated the light in this room). Although the lengths of the exchanged packets can be different, features, such as TCP flags, \textcolor{black}{destination ports} and source ports can be common depending on the device and event type. The sequence of packets and \textcolor{black}{their} features remain the same each time a particular event occurs. As a result, such packet bursts can be used to detect that an event has occurred at a particular time. 


The \sol controller keeps a log file for each device based on a chronological occurrence of events as shown \textcolor{black}{in Figure \ref{fig:Interaction1}(c). When} an event is detected, the controller updates the log file of the device that generated this event with information related to the event. The controller stores logs related to an event $e$, at least, until $e$ and all events triggered due to $e$ have been validated. After that, logs of validated events can be archived on a remote cloud and be deleted by the controller if it does not have adequate resources.


Prior research \cite{10.1007/978-981-10-5421-1_15, https://doi.org/10.48550/arxiv.1907.11797, 7013127} has shown that each event has a unique pattern of exchanged packets depending on the device and its manufacturer, which can be used as a signature. This is due to the fact that IoT devices typically conduct a limited number of operations (\ie space of potential events). In \sol, we create packet signatures based on tuples of different header fields. Specifically, we have taken into account the following features: source and destination port numbers, the exchanged packet lengths, TCP flag types, the time span of each event, and the direction of the packets within that time span. Based on these features, \sol generates packet signatures for various events.







\subsection{Packet Level Anomaly Detection}

An event will be considered as a packet-level anomaly if the set of exchanged packets for a particular event does not match (or does not have a similar pattern) to any prior generated event signatures. An anomaly can occur for various reasons, such as faulty devices and/or sensors, and device(s) compromised by an attacker.

\sol uses a light-weight machine learning model running on the controller to detect packet-level anomalies. The model is trained based on previously extracted event signatures. 
Since each IoT device typically performs a specific set of operations over time, the features of potential events for available IoT devices can be collected as part of the training dataset. Once the model is trained, the controller uses this model to identify whether incoming events are valid. If an incoming event is valid based on the prediction of the model, the controller allows the event to continue. Otherwise, the event is deemed \textcolor{black}{an anomaly and the controller terminates it.}



\subsection{Device Interaction Validation} 


If no packet-level anomalies are identified for an event, the controller will let the event be executed. The controller will also log event-related information, such as which device triggered the event and which device’s state has been changed because of that particular event. Logs of event-related information will be used to validate events that occur over time based on conditions defined by users and/or the vendor of each device. We provide examples of such conditions below.


We define a device interaction as an event on a device triggered by another device in the IoT environment. Similarly, the same event can be triggered \textcolor{black}{by events of different devices.} For example, an event that turns on a smart bulb can be triggered by a motion detector. The same bulb can also be turned on by a light sensor. An interaction among devices will be considered as a valid one if it follows user- or vendor-defined conditions, so that the IoT environment as a whole operates as expected. For instance, a user can set a condition that if the room temperature as detected by a smart temperature monitor exceeds a certain threshold value, then the monitor will send a command via the controller to turn \textcolor{black}{on the AC and the fan.} In this case, turning on the AC and the fan by the temperature monitor will be considered as valid interactions. However, if the temperature monitor tries to turn on a light, \textcolor{black}{this will be considered as an invalid interaction (anomaly)}.

\begin{figure*}
  \includegraphics[width=\textwidth]{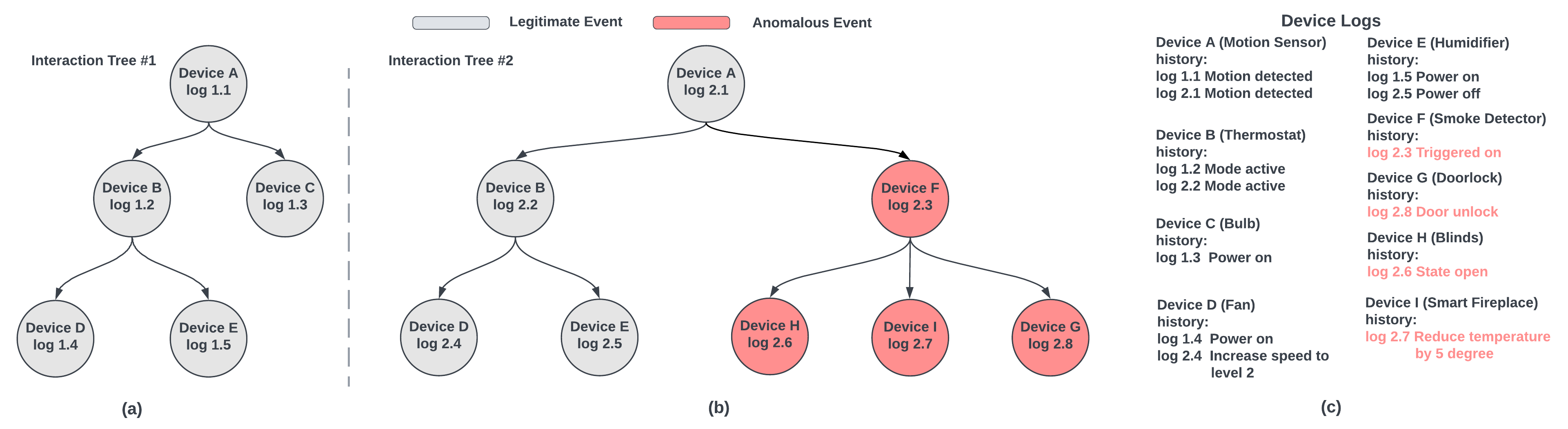}
\caption{Illustration of interactions among IoT devices.}
\label{fig:Interaction1}
\vspace{-0.2cm}
\end{figure*}

The controller creates and maintains data structures over time, called interaction trees. These trees represent sequences of interconnected events triggered by a measurement or a reading of a device. \textcolor{black}{The device that starts an interaction is the root device of a tree.} For example, in Figure~\ref{fig:Interaction1}(a), \textcolor{black}{device A is the} root device of the interaction tree. When the root device generates a new measurement/reading, this triggers the creation of a new interaction tree. Figure \ref{fig:Interaction1}(a) shows how interaction trees are formed in \sol. For interaction tree 1 of device A, two events are triggered \textcolor{black}{on two different} devices (device B and device C). Furthermore, the event of device B further triggers events \textcolor{black}{on device} D and device E.

\begin{figure}[th]
\centering
\vspace{-0.3cm}
\includegraphics[width=0.45\linewidth]{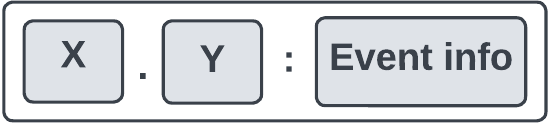}
\caption{Device log \textcolor{black}{generation in \sol. ``X'' is the sequence number of a measurement/reading of a root device, while ``Y'' is the  sequence number of a specific event within the ``X'' interaction tree.}}
\vspace{-0.3cm}
\label{fig:log}
\vspace{-0.3cm}
\end{figure}

To record the device interactions and the corresponding measurements of root devices in a consistent manner, we introduce a key-value pair-based logging format, which is used to map the information of each event to a unique key. A unique key is generated based on two different IDs as shown in Figure \ref{fig:log}. The first part of the key ``X'' is the sequence number of the measurement/reading of a root device (for every new interaction tree, the sequence number is incremented) and ``Y'' is the sequence number of a specific event within the ``X'' interaction tree. We elected to not use timestamps as a part of the unique key generation, since time synchronization among IoT devices is a challenge on its own.

Within an IoT environment, each IoT device may act as the root device \textcolor{black}{of different interaction trees.} Each tree is generated dynamically when the root device generates a new measurement/reading (\eg a temperature sensor may capture a new reading periodically every few seconds). \sol makes use of automation rules and conditions (\eg defined by users or device manufacturers) to verify interactions among devices. For example, as shown in Figure~\ref{fig:Interaction1}(b), a measurement/reading captured by device A triggers multiple events on other IoT devices. In this example, an anomalous event has occurred on device F, and \sol detects the anomaly based on the automation rules and conditions. Subsequently, device F interacted with several other devices, as illustrated in Figure~\ref{fig:Interaction1}(b). As a result, due to the anomalous event on device F, the state of device G, device H, and device I will also be affected.

\subsection{Rollback}

To mitigate the effects of anomalies that propagate among devices and help affected devices revert their state to the last known stable state, \sol offers an automated recovery mechanism, \textcolor{black}{called ``rollback''.} The rollback process begins as soon as an interaction anomaly is detected. At this point, the \sol controller will use the corresponding interaction tree to identify the devices that got affected due to this interaction anomaly. For instance, in Figure \ref{fig:Interaction1}(b), an anomalous \textcolor{black}{event on device F causes anomalous events on devices H, I, and G.} \sol creates a list of those devices and analyzes the logs of these devices to find out the last known stable state for each device. For instance, device F was triggered on due to the anomalous interaction with device A. Therefore, the last stable state of device F was ``triggered off''. Hence, \sol sends commands to affected devices, so that they roll back their state to the last known stable state (in this example, ``triggered off'' for device F). Finally, \sol isolates the device that initiated the anomaly until the device owner or administrator can troubleshoot.

The above mechanism essentially assumes that all interactions are legitimate until an anomaly is found. In other words, \sol allows interactions to take place among devices (and potential anomalies to propagate among devices), while verifying such interactions asynchronously. If an anomaly is detected, \textcolor{black}{the state of all affected devices} will be rolled back. An alternative to this mechanism would be to first verify each interaction that a device would like to perform and, only if this interaction is valid, let the device interact with another device. In this case, anomalies would not be able to propagate, since devices would interact with each other only after each interaction has been verified. {\revone However, verifying automation rules, especially when the number of IoT devices grows and/or the rules are complex, may not happen instantly. To this end, we would need to essentially stop the operation of the IoT environment until an interaction has been verified.}

\section{Evaluation}
\label{sec:eval}

In this section, we first present our evaluation setup and we then discuss our evaluation results.




\begin{table*}[]
\begin{minipage}{1\textwidth}
\centering
\caption{Datasets used for the evaluation of \sol. }
\vspace{-0.2cm}
\label{tab: datasets}
\resizebox{0.95\linewidth}{!}{
\begin{tabular}{|c|c|c|c|c|c|}
\hline
{\textbf{Dataset}} & {  \textbf{Setup}} & {  \textbf{No. of devices}} & {  \textbf{Duration}} & {  \textbf{Content Type}}    & {  \textbf{Data Type}} \\ \hline
{  MUD~\cite{10.1145/3314148.3314352}}       & {  Smart Home}     & {  10}                      & {  21 days}           & {Packet level Data + Device Interaction Data} & {  pcap}               \\ \hline
{  Mon(IoT)r~\cite{10.1145/3355369.3355577}}  & {  Smart Home}     & {  45}                      & {  3 days per device}    & {Packet level Data + Device Interaction Data}               & {  pcap}               \\ \hline
{  PingPong~\cite{8819458}}  & {  Smart Home}     & {  16}                      & {  15 days}           & {Packet level Data + Device Interaction Data}               & {  pcap}               \\ \hline
{  Testbed data}     & {  Smart Home}     & {  12}                      & {  10 days}           & {Packet level Data + Device Interaction Data} & {  pcap}               \\ \hline
\end{tabular}
}
\end{minipage}
\end{table*}

\subsection{Evaluation Setup}

\textcolor{black}{Our evaluation process involves three parts: (i) we first use open-source IoT device datasets; (ii) we then build a small-scale testbed of IoT devices, which we use to deploy \sol in the real world; and (iii) we finally compare \sol to relevant approaches that have been previously proposed.}




\subsubsection{Open-Source IoT Datasets} In this part of our evaluation, we made use of three publicly available datasets of IoT device data. These datasets were collected through controller platforms, such as IFTTT and SmartThings. Based on this data, we generated packet-level signatures and then replayed this data in the \sol controller for packet-level anomaly detection and device interaction validation. 
We summarize the characteristics of these datasets in Table~\ref{tab: datasets} and we further discuss them below.


\noindent \textbf{Manufacturer Usage Description (MUD) dataset~\cite{10.1145/3314148.3314352}:} This dataset includes real-world traffic traces of 10 different types of smart devices collected over a period of 21 days. It contains two types of traces: volumetric attack traces (\eg ARP spoofing, TCP/UDP flooding) and benign traces. This dataset provides information about the number of MUD flows per minute, the start and the end time of an attack, and the MUD flows impacted due to an attack.  

\noindent \textbf{Mon(IoT)r dataset~\cite{10.1145/3355369.3355577}:} The Mon(IoT)r dataset was created by collecting traces from a number of devices \textcolor{black}{over 85 days.} The data was collected in two different countries (in the US and the UK). Each PCAP file of the dataset represents the traces related to an event of a specific device. The dataset has multiple instances of the same event for a device, which {  helps identify} the patterns (generate signatures) for each event. This dataset also contains the timestamps of each event.    

\noindent \textbf{PingPong dataset~\cite{8819458}:} This dataset contains PCAP files with network traffic traces of 22 popular commercially available IoT devices. \textcolor{black}{The creators of the dataset} identified different functions or events of these devices and captured them in these PCAP files. The dataset also includes the timestamps of the captured functions for each device. 

\subsubsection{Small-Scale Smart Home Testbed} 

\begin{figure}
\centering
\includegraphics[width=1\columnwidth]{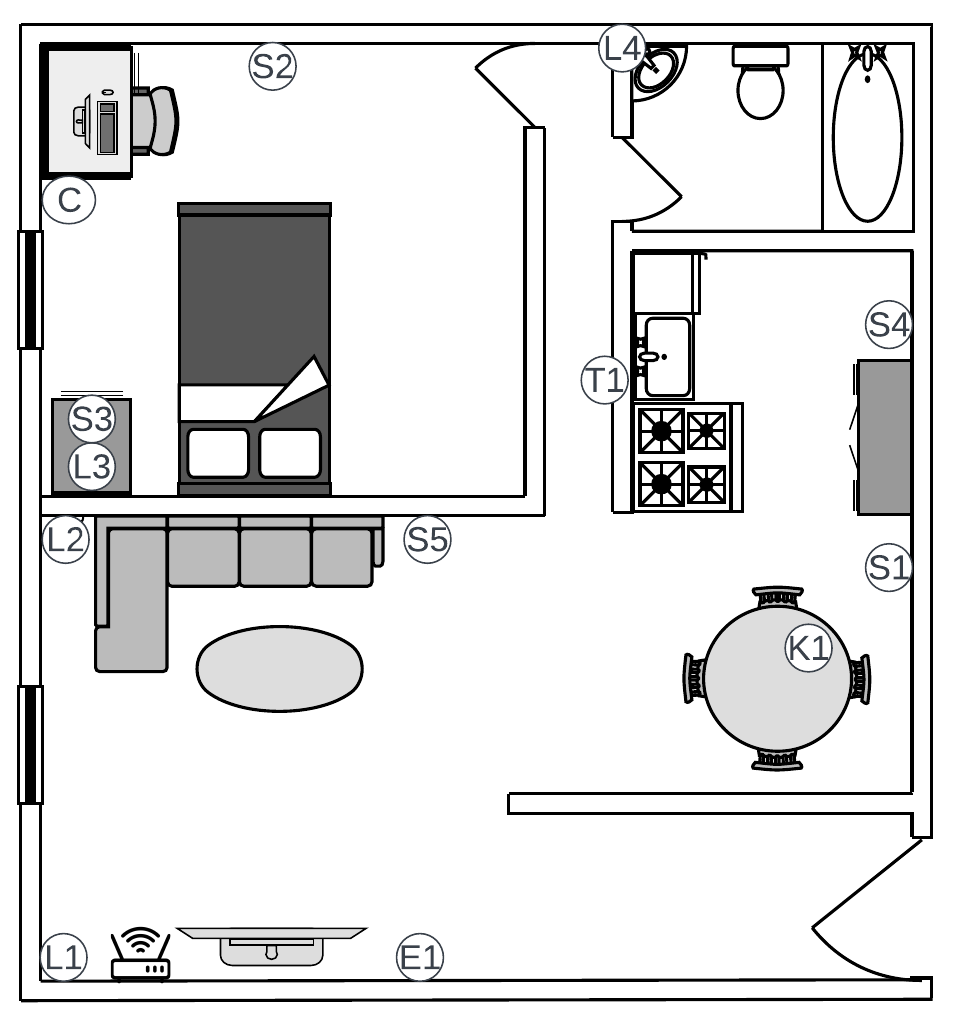}
\caption{Floor plan of our IoT testbed and device deployment layout.}
\label{fig:testbed}
\end{figure}

We retrofitted a research space to an one-bedroom apartment \textcolor{black}{to create a} small-scale smart home testbed, where we evaluated the \sol design.  Figure~\ref{fig:testbed} shows the layout of the testbed and the locations of the used IoT devices. Table~\ref{tab:device-list} lists the details for the devices we used. Our testbed contains a total of 12 IoT devices. These devices can interact with each other via the controller of \sol and trigger different events based on predefined conditions. All devices are wirelessly connected to the controller \textcolor{black}{over a local network.} 

We used an Intel Next Unit Computing (NUC) mini kit {  as an} \sol controller in our testbed. The controller runs a daemon program that can monitor all packets exchanged among the IoT devices. We used PyShark, a popular Python library to capture network traffic and interaction traces from the devices over a period of 10 days. We also used different APIs that come from device vendors to interact with the devices and collect the timestamps of various device functions. \textcolor{black}{More information about our testbed dataset is provided in Table~\ref{tab: datasets}.} 
We make the dataset collected from our testbed and the \sol implementation code publicly available for the community to use\footnote{Our testbed dataset and \sol implementation code are available at \url{https://github.com/Hasniuj-Zahan/IoT-anomaly-detector}.}.

\begin{table}[]
\caption{List of IoT devices used in our testbed and their abbreviated labels.}
\centering
\label{tab:device-list}
\begin{tabular}{|l|l|}
\hline
\textbf{Abbr} & \textbf{Device Name}               \\ \hline
C             & Intel NUC mini kit as controller   \\ \hline
L1            & Govee Smart Light                  \\ \hline
L2            & Sengled Smart Bulb                 \\ \hline
L3            & Tp-link Smart Bulb                 \\ \hline
L4            & Kasa Smart Light                   \\ \hline
S1            & TP-link Smart Plug                 \\ \hline
S2            & Teckin Smart Plug                  \\ \hline
S3            & WeMo Smart plug                    \\ \hline
S4            & KMC 4 Outlet WiFi Smart Plug       \\ \hline
S5            & Amazon Smart Plug                  \\ \hline
T1            & Ecobee Thermostat                  \\ \hline
E1            & 4th Generation Echo Dot with Clock \\ \hline
K1            & Korex Smart Kettle                 \\ \hline
\end{tabular}
\end{table}

\textcolor{black}{\subsubsection{Comparison to previously proposed approaches} We compared the \sol design to an Artificial Neural Network (ANN) based framework for IoT intrusion and anomaly detection \cite{hodo2016threat} and a Convolutional Neural Network (CNN) based framework for IoT anomaly and intrusion detection \cite{li2020robust}. For {  this} comparison, we used the four datasets mentioned above (MUD, Mon(IoT)r, PingPong, and {  our} testbed dataset).} 

\subsubsection{Evaluation Metrics}
For the evaluation of the detection of packet-level anomalies, \textcolor{black}{we used four machine} learning algorithms: a random forest algorithm~\cite{Biau_Scornet_2016}, a k-Nearest Neighbors (kNN) algorithm~\cite{altman1992introduction}, a decision tree~\cite{song2015decision}, \textcolor{black}{ and an autoencoder \cite{hinton1993autoencoders}}. We used these algorithms instead of more complex models (\eg a neural network~\cite{dayhoff1990neural}) due to their lightweight nature. For the validation of device interactions, we implemented the \sol controller in software (based on the design described in Section \ref{sec:design}). We consider the following metrics for our evaluation:


\begin{enumerate} [wide, labelwidth=!, labelindent=0pt, nosep]

\item {\em Inference time:} The time \textcolor{black}{needed for the} \sol controller to identify whether: (i) a specific event contains packet-level anomalies; or (ii) a device interaction is illegitimate (anomalous). 


\item{\em Memory usage:} The amount of memory required by the \sol controller to detect packet-level anomalies and validate interactions.  

\item{\em CPU usage:} The CPU usage of the \sol controller while validating the interactions among devices and identifying packet-level anomalies.

\item{\em Accuracy:} The accuracy of different models in terms of identifying packet-level anomalies. 

\item{\textcolor{black}{\em Precision}:} \textcolor{black}{The precision of a model refers to the ratio of the number of true positive anomaly predictions to the total number of positive anomaly predictions.} 

\item{\textcolor{black}{\em F1 score}:} \textcolor{black}{The F1 score of a model indicates the balance between the precision and the recall of the model for the identification of anomalies.}

\end{enumerate}

\subsection{Evaluation Results}


\subsubsection{Evaluation of packet-level anomaly detection}

\begin{figure}
\centering
\includegraphics[width=1\linewidth]{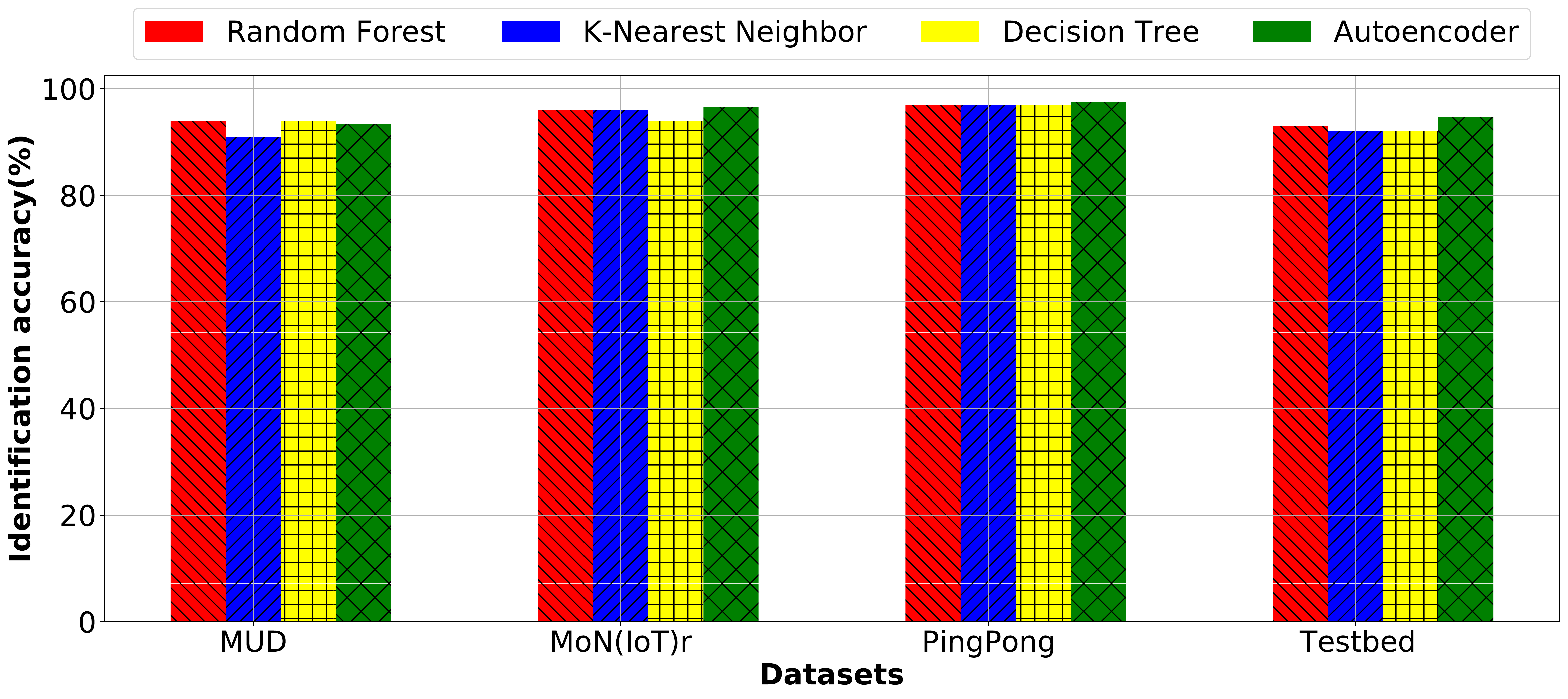}
\caption{Packet-level anomaly detection accuracy using \sol for different datasets.}
\label{fig:accuracy}
\end{figure}

\noindent\textbf{Packet-level anomaly detection accuracy:} 
In Figure \ref{fig:accuracy}, we present our results on the identification accuracy of packet-level anomalies for different datasets and learning algorithms. 
\textcolor{black}{Our results demonstrate that through the use of packet-level {  signatures} that are created for IoT devices, \sol is able to achieve over 90\% and, in some instances, up to 98\% of accuracy.}

\begin{figure}
\centering
\includegraphics[width=1\columnwidth]{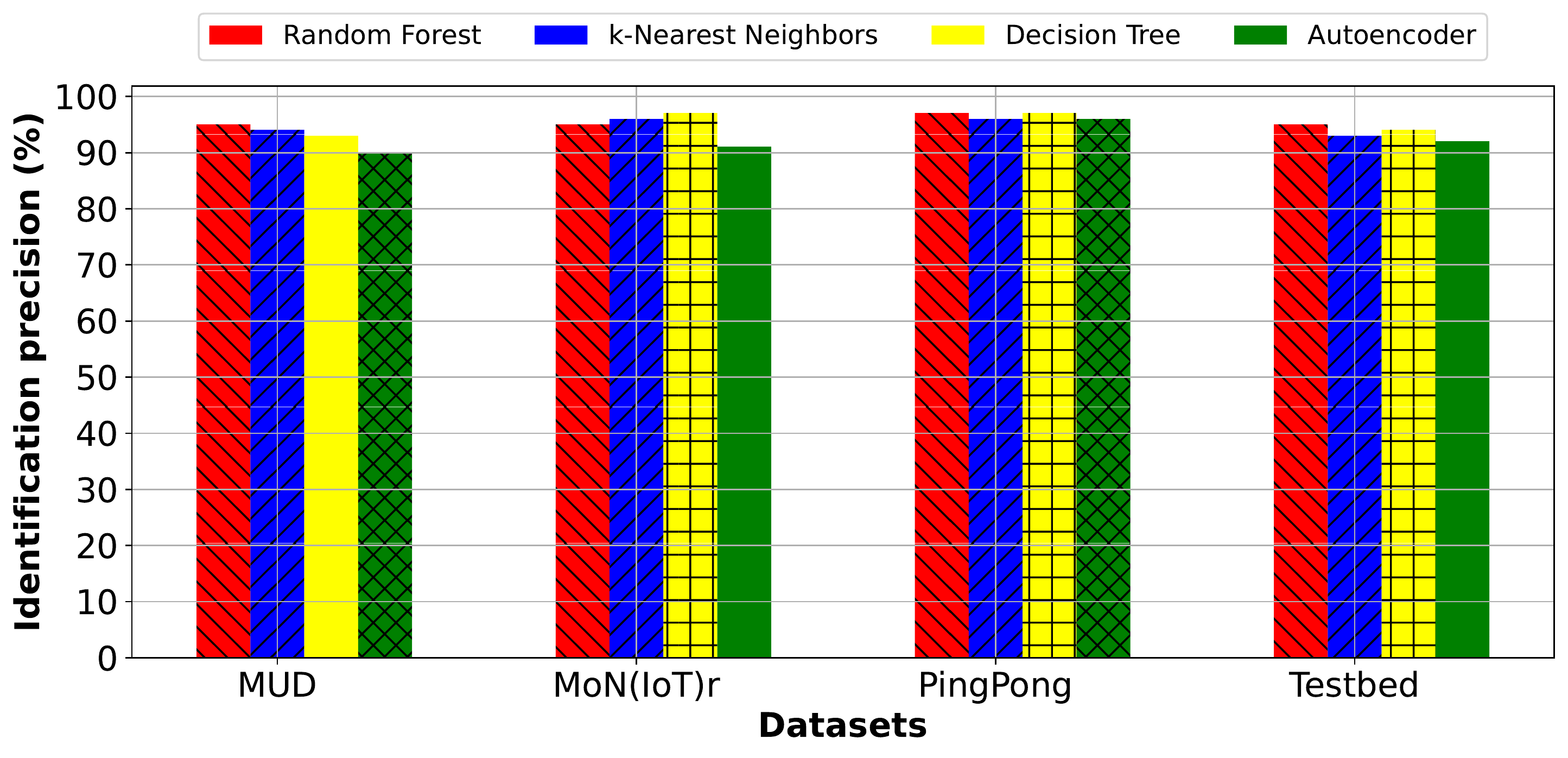}
\caption{\textcolor{black}{Packet-level anomaly detection precision score using \sol for different datasets.}}
\label{fig: PrecisionPacketanomaly}
\end{figure}

\noindent\textcolor{black}{\textbf{Precision of packet-level anomaly detection:} Figure \ref{fig: PrecisionPacketanomaly} shows the precision results of packet-level anomaly identification for different models and datasets. Similar to the accuracy results, \sol achieves precision scores over 90\% in all cases.}

\begin{figure}
\centering
\includegraphics[width=1\linewidth]{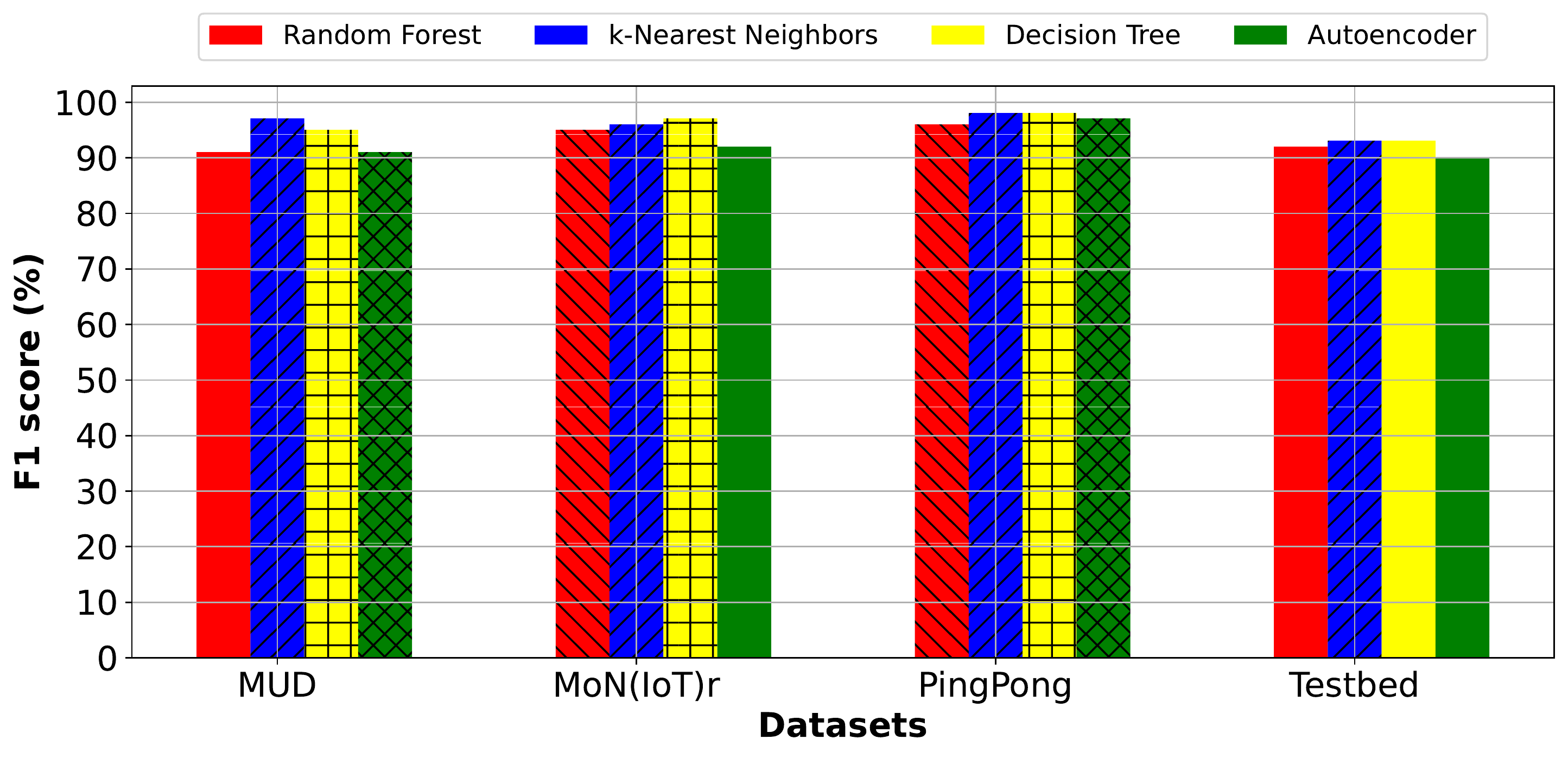}
\caption{\textcolor{black}{Packet-level anomaly detection F1 score using \sol for different datasets.}}
\label{fig:F1score}
\end{figure}

\noindent\textcolor{black}{\textbf{F1 score of packet-level anomaly detection:} In Figure \ref{fig:F1score}, we present the {  F1 scores} of different anomaly identification models and datasets. The results of Figure \ref{fig:F1score} indicate that all models can identify packet-level anomalies with low false positive and false negative rates. Specifically, the models achieve F1 scores over 90\% (and up to 98\% in some instances). }

\begin{figure}
\centering
\includegraphics[width=1\columnwidth]{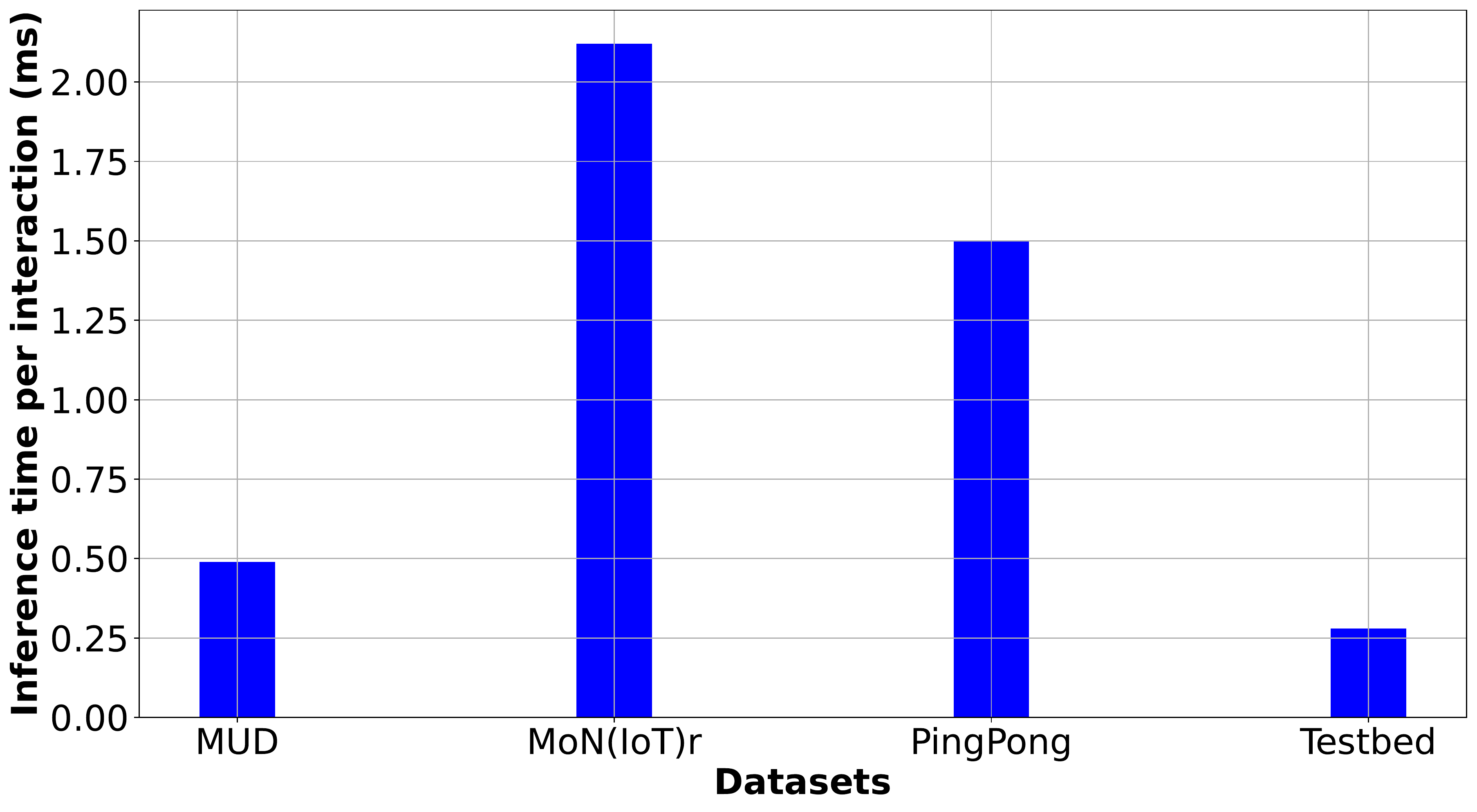}
\caption{\textcolor{black}{Inference time for packet-level anomalies.}}
\label{fig: InferenceTime}
\end{figure}

\noindent\textbf{Inference time:} In \textcolor{black}{Figure \ref{fig: InferenceTime}}, we present the inference time per packet-level anomaly. 
Our results indicate that \sol requires less than 2.12 milliseconds on average to identify whether an event contains a packet-level anomaly. The inference time depends on the nature of each device and dataset. For example, in the MoN(IoT)r dataset, the average inference time is higher as compared to other datasets. This is due to the fact that the IoT devices in this dataset are more complex in nature (\eg fireTV, smart fridges), thus their events are represented through more complex communication protocols that involve more packet exchanges than rudimentary devices (\eg smart bulb, smart switch). Further analysis of our results shows that another factor, which can impact inference time, is how ``noisy'' communication between devices is. In this context, noise typically includes traffic that is exchanged between devices, but is not related to actual events (\eg traffic related to miscellaneous network protocols or traffic related to firmware updates).

\begin{figure}
\centering
\includegraphics[width=1\columnwidth]{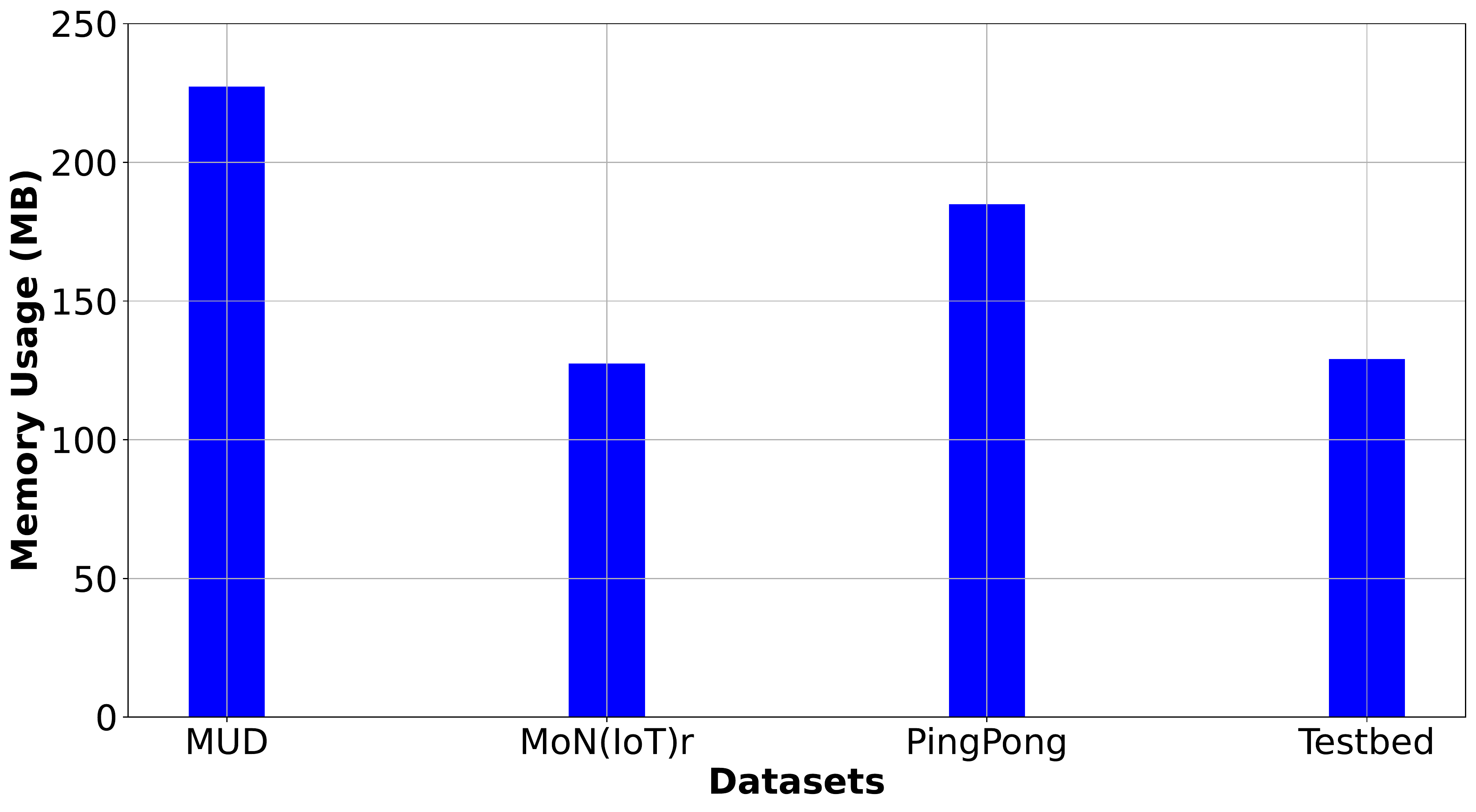}
\caption{\textcolor{black}{Memory usage during packet-level anomaly detection.}}
\label{fig: MemoryUsagePacketanomaly}
\end{figure}

\noindent\textbf{Memory usage:} In {  Figure} \ref{fig: MemoryUsagePacketanomaly}, we present the results of the controller's memory usage while detecting packet-level anomalies. Our results indicate that the controller requires a few hundred MB (up to 227MB) for the identification of packet-level anomalies. To this end, \sol can detect packet-level anomalies even when deployed on controller devices with limited memory resources.

\begin{figure}
\centering
\includegraphics[width=1\columnwidth]{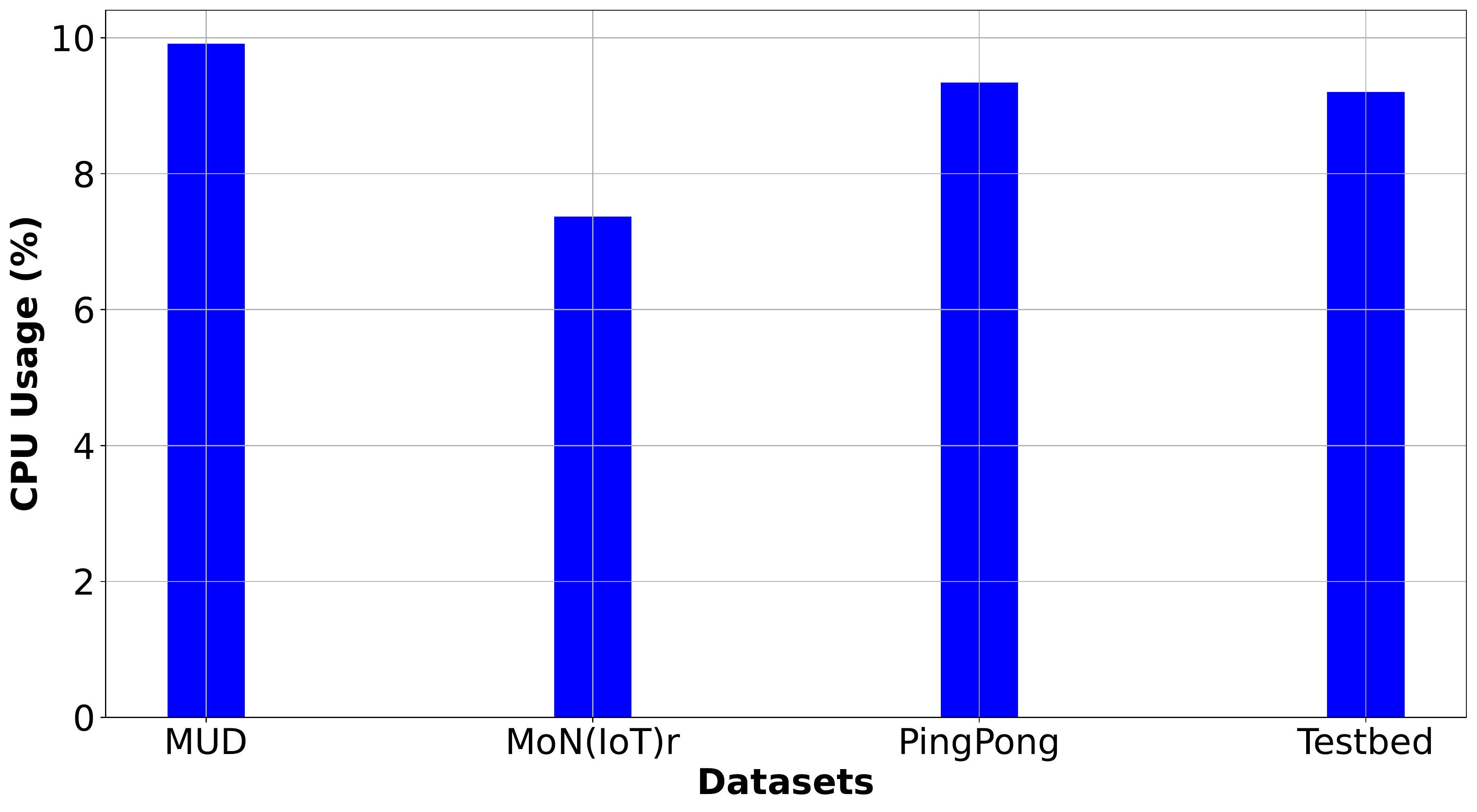}
\caption{\textcolor{black}{CPU usage during packet level anomaly detection.}}
\label{fig: CPUUsagePacketanomaly}
\end{figure}

\noindent\textbf{CPU usage:} In {  Figure} \ref{fig: CPUUsagePacketanomaly}, we present the CPU load of the \sol controller during the detection of packet-level anomalies. The controller uses less than 10\% of its available CPU in order to identify packet-level anomalies in the traffic exchanged between IoT devices. Our results demonstrate that \sol can detect packet-level anomalies even when the deployed controller has limited CPU resources.

\subsubsection{Evaluation of interaction validation and rollback}

\begin{figure}
\centering
\includegraphics[width=1\columnwidth]{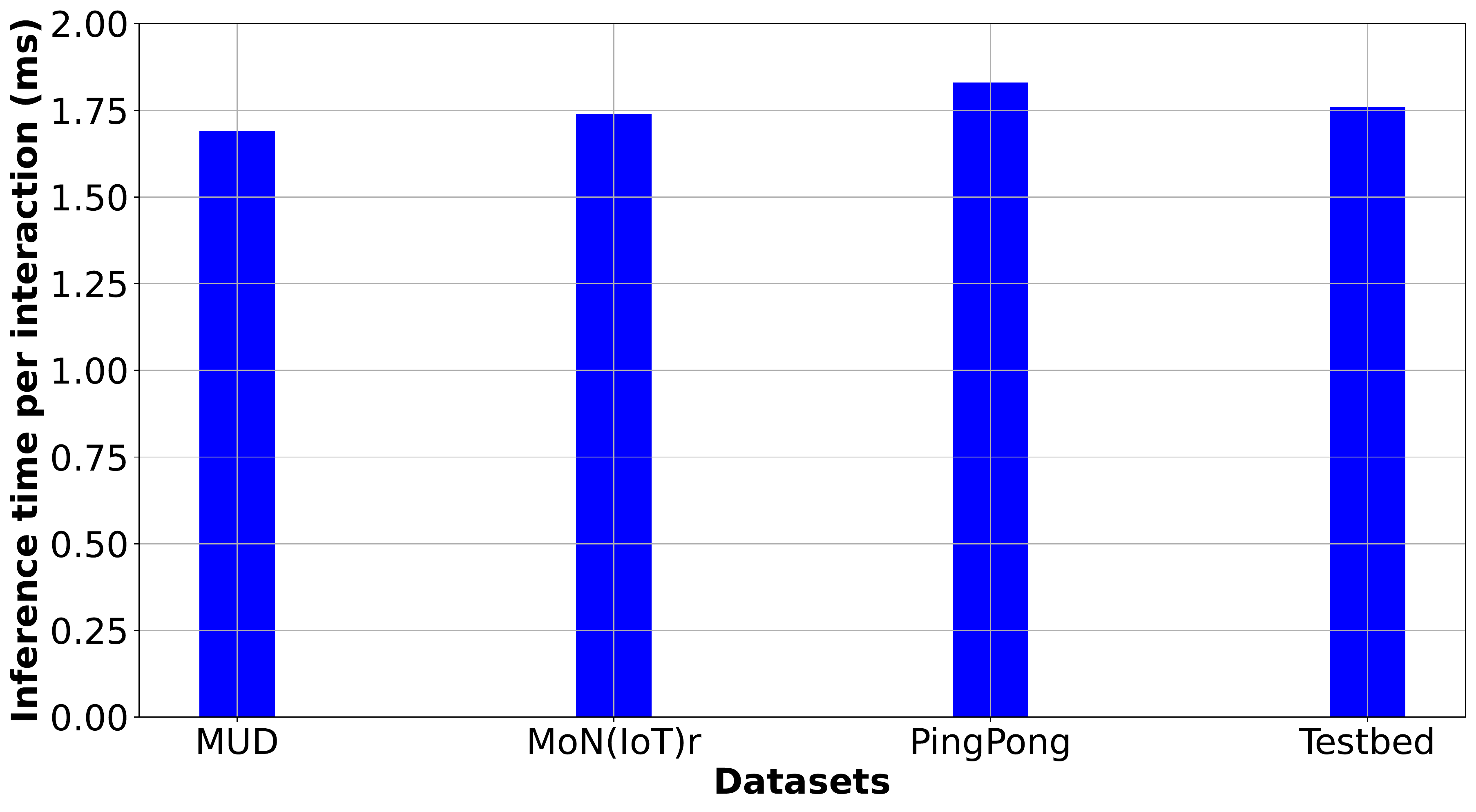}
\caption{\textcolor{black}{Time for inference and rollback per device interaction.}}
\label{fig: InteractionInferenceTime}
\end{figure}

\hfill \break
\noindent\textbf{Inference time:} In \textcolor{black}{Figure \ref{fig: InteractionInferenceTime}}, we present the inference time per device interaction based on different datasets. Our evaluation results show that the \sol controller requires 1.74ms-1.83ms to validate an interaction and roll back to the last known stable state, if there is an anomaly. 

\begin{figure}
\centering
\includegraphics[width=1\columnwidth]{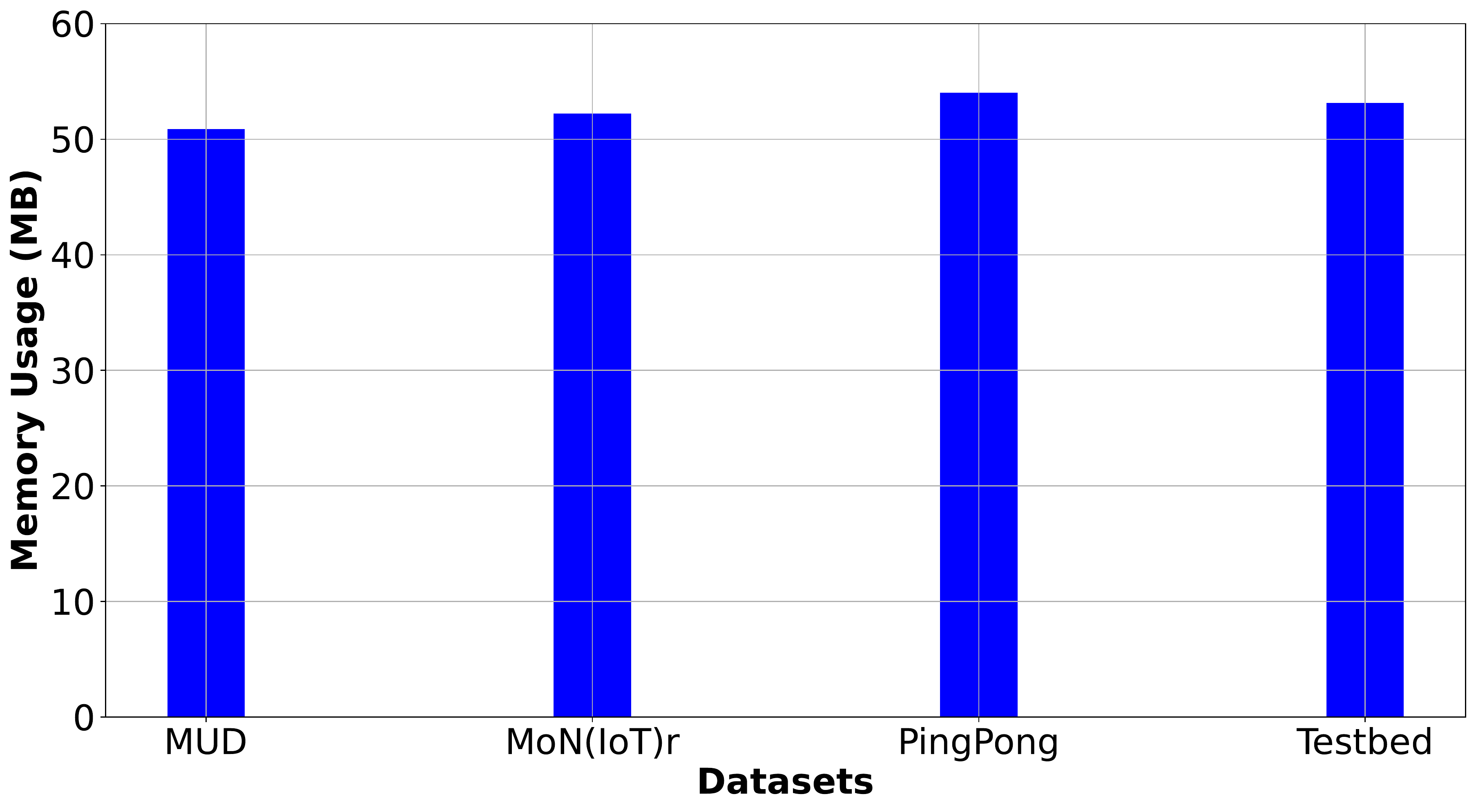}
\caption{\textcolor{black}{Memory usage during device interaction validation and rollback.}}
\label{fig: MemoryUsageInteractions}
\end{figure}

\noindent\textbf{Memory usage:} \textcolor{black}{Figure \ref{fig: MemoryUsageInteractions}} shows the required memory to validate device interactions for different datasets. To run the device interaction validation module of the \sol~{  controller}, it takes around 54 MB of memory on average to validate an interaction among devices. This module not only detects {  interactions} among devices, but also detects device interaction anomalies and {  performs rollback, if necessary.}

\begin{figure}
\centering
\includegraphics[width=1\columnwidth]{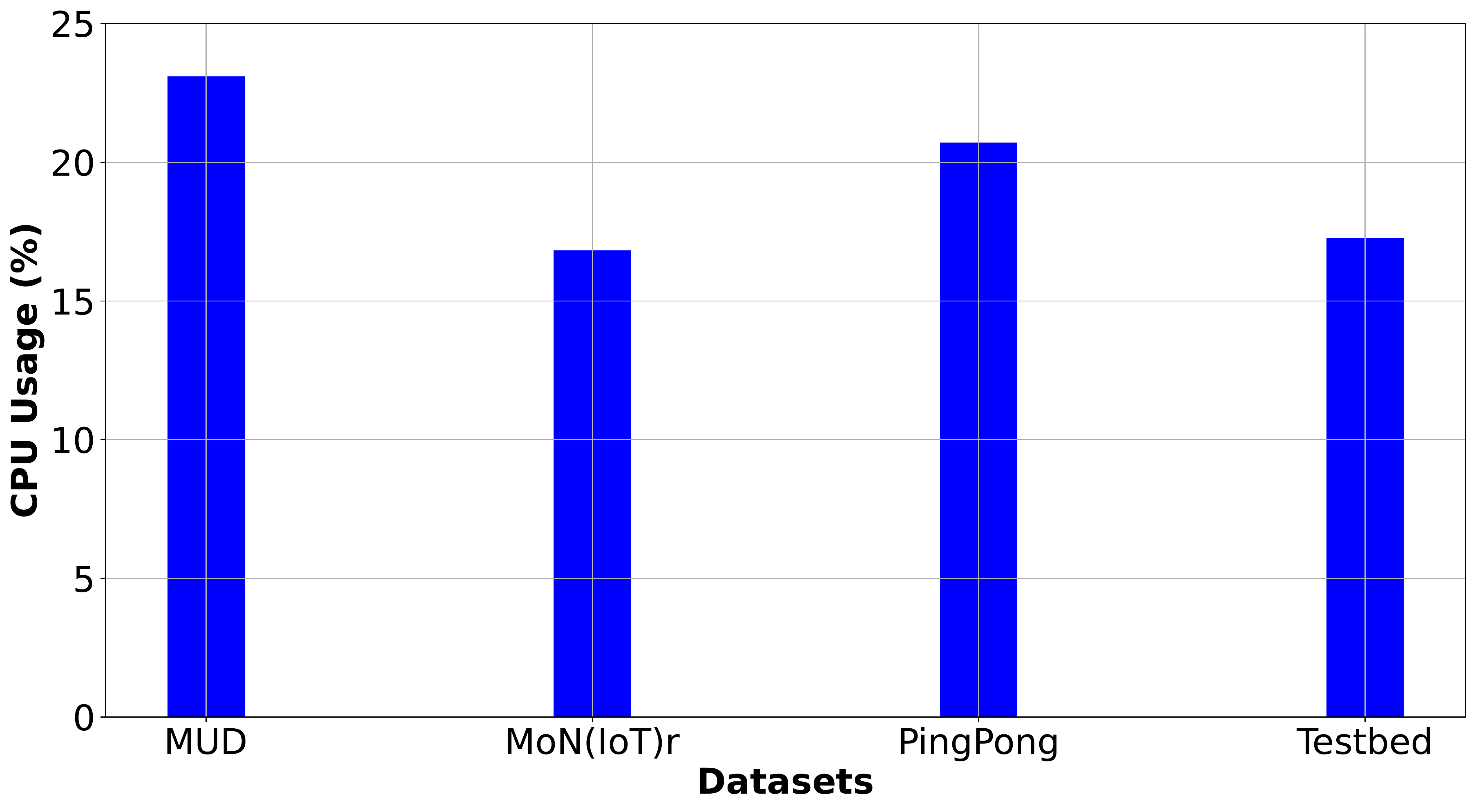}
\caption{\textcolor{black}{CPU usage during device interaction validation and rollback.}}
\label{fig: CPUUsageInteractions}
\end{figure}

\noindent\textbf{CPU usage:} \textcolor{black}{Figure \ref{fig: CPUUsageInteractions}} shows the results of CPU usage {  during device interaction validation and rollback when} an anomaly is detected. 
{  The CPU usage during the} device interaction validation and rollback process is 16.8\%-23.1\%. The results show that {  these processes are lightweight, so that they can run on IoT controllers} with limited computing {  resources.} 

\subsubsection{\textcolor{black}{Comparison to prior related frameworks}}

\begin{figure}
\centering
\includegraphics[width=1\linewidth]{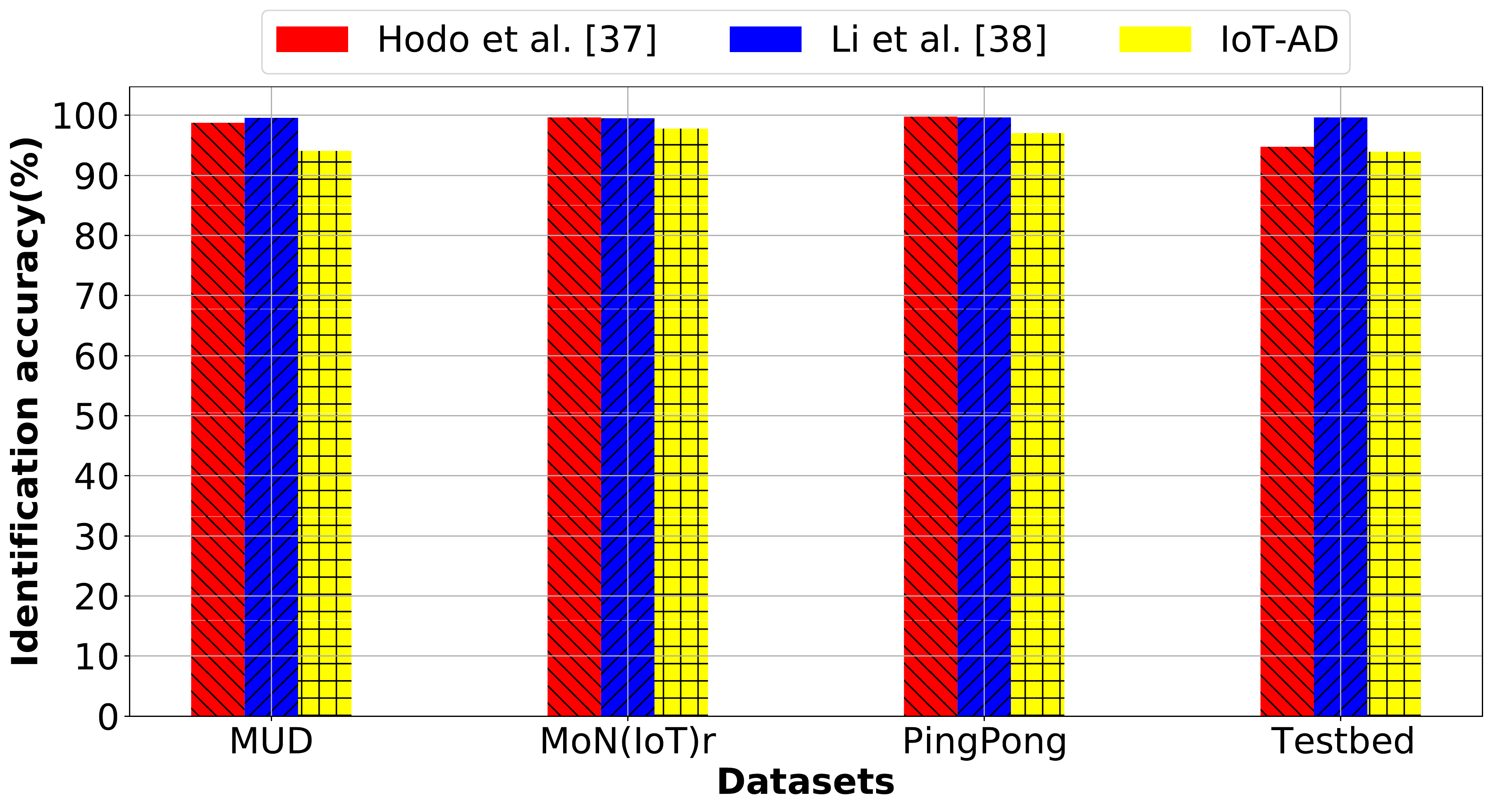}
\caption{\textcolor{black}{Comparison of packet-level anomaly detection accuracy between \sol and other frameworks.}}
\label{fig:comparison_accuracy}
\end{figure}

\begin{figure}
\centering
\includegraphics[width=1\linewidth]{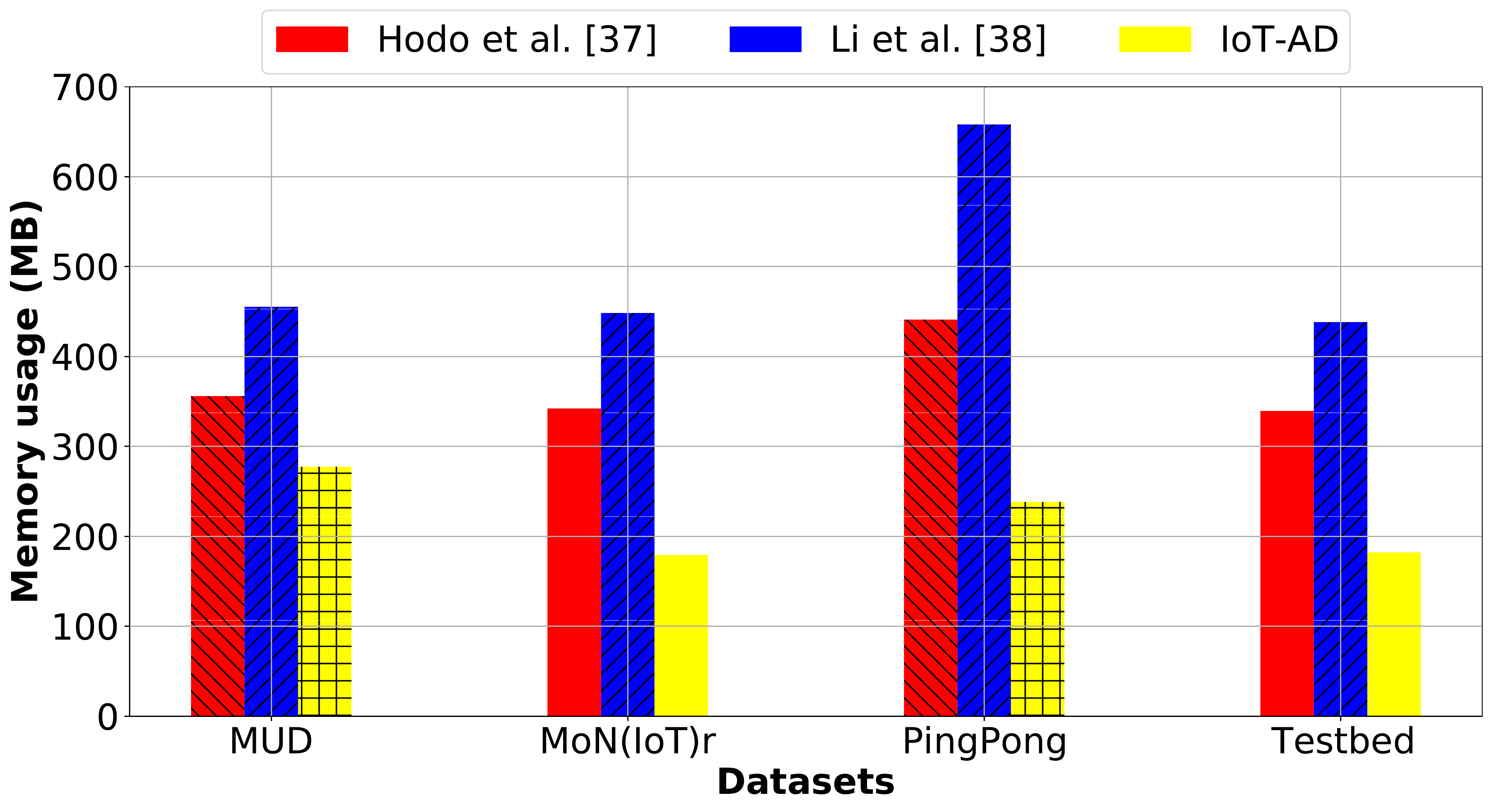}
\caption{\textcolor{black}{Comparison of memory usage between \sol and other frameworks.}}
\label{fig:comparison_memory}
\end{figure}

\begin{figure}
\centering
\includegraphics[width=1\linewidth]{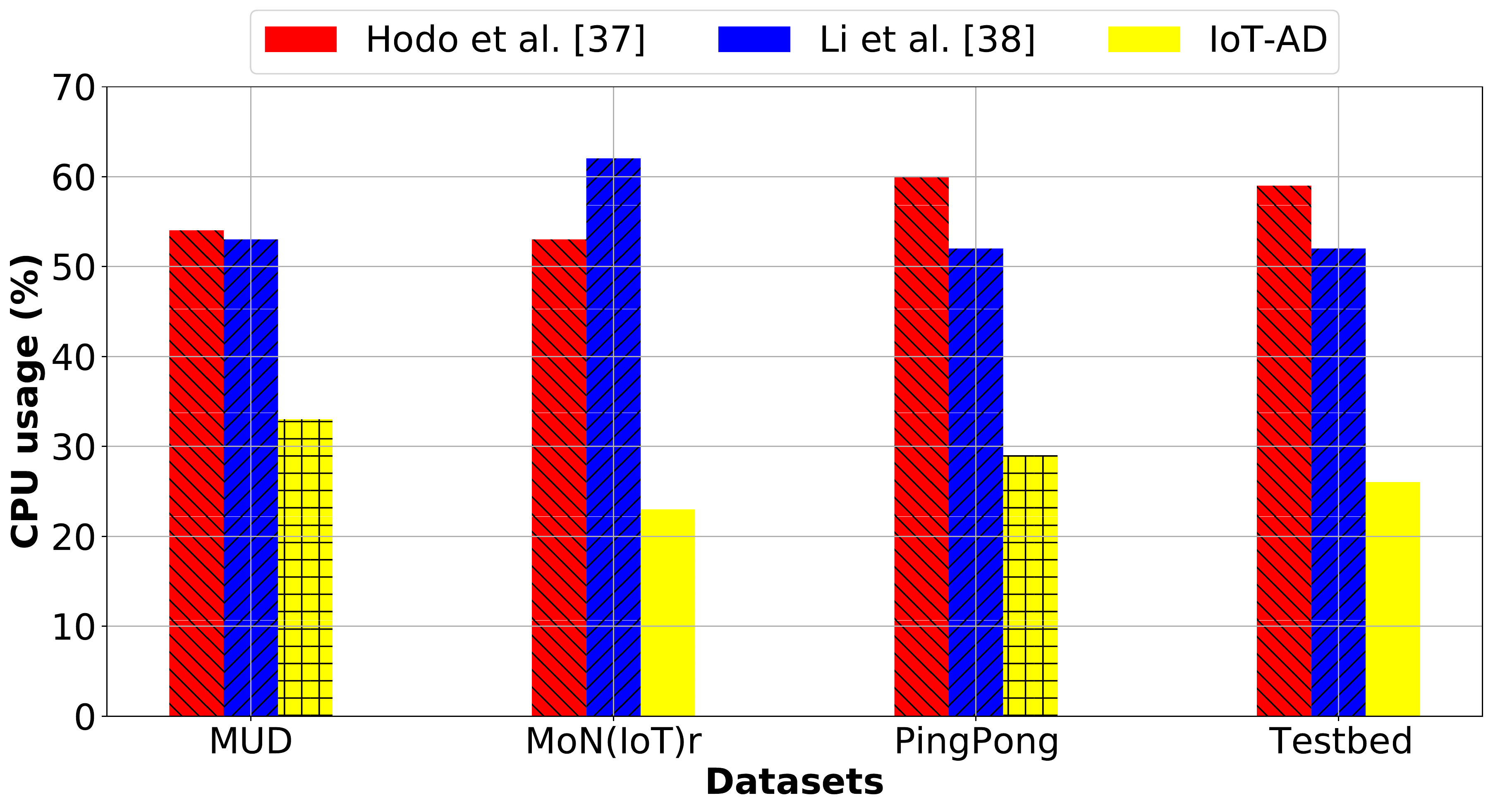}
\caption{\textcolor{black}{Comparison of CPU usage between \sol and other frameworks.}}
\label{fig:comparison_cpu}
\end{figure}

\begin{figure}
\centering
\includegraphics[width=1\linewidth]{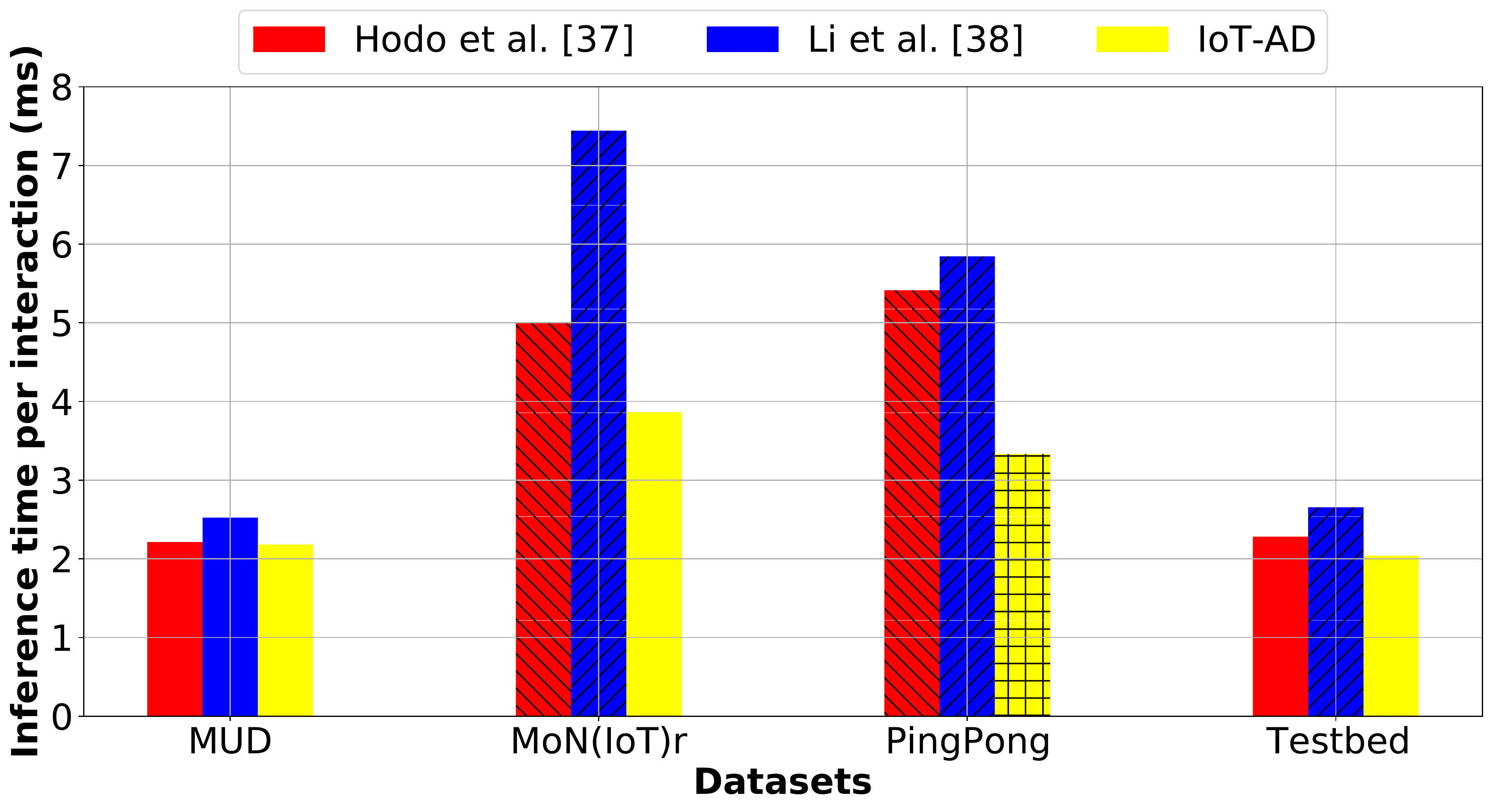}
\caption{\textcolor{black}{Comparison of inference time between \sol and other frameworks.}}
\label{fig:comparison_inference}
\end{figure}

\textcolor{black}{Although \sol achieves 1\%-3\% lower anomaly detection accuracy as compared to ANN and CNN based frameworks (because of their complex architectures) as shown in Figure \ref{fig:comparison_accuracy}, Figures \ref{fig:comparison_memory} and \ref{fig:comparison_cpu} indicate that \sol results in 22\%-63\% lower memory usage and 39\%-62\% lower CPU usage. These results indeed verify the {  lightweight} nature of \sol, which makes it suitable for deployment on resource-constrained IoT device controllers. In addition, Figure \ref{fig:comparison_inference} shows that the inference time of \sol is up to 48\% lower than other frameworks, which is desirable for the detection of anomalies in real-time. Finally, unlike other frameworks, \sol provides mechanisms to detect packet-level anomalies, validate interactions among interconnected and interdependent IoT devices, and rollback to the most recent known stable state(s) in case that an anomaly is detected.}

\section{Discussion}
\label{sec:discussion}

\textcolor{black}{In this section, we discuss the process of selecting a device that can act as the \sol controller and the placement of this controller. We also provide a security analysis of \sol and discuss the limitations of the \sol design as a whole.}

\subsection{\sol Controller Selection and Placement}

During the evaluation of the \sol design, we chose a smart hub to act as the controller in the IoT environment. Alternatively, we could have chosen an IoT device with adequate resources to act as the controller (\eg a smart TV, a smart refrigerator). As a rule of thumb, any device with adequate resources to identify packet-level and interaction anomalies (such resource requirements have been quantified through our evaluation process in Section~\ref{sec:eval}) should be able to act as the controller.

The \sol controller could be deployed within a local IoT environment or it could be remotely deployed on a cloud. A local controller could reduce the overall latency when it comes to detecting anomalies and helping IoT devices interact with each other. Another benefit of having a local controller is that an external network observer (\eg an Internet Service Provider (ISP) or a cloud service provider) cannot analyze the traffic to identify what is happening inside an IoT environment \cite{DBLP:journals/corr/ApthorpeRF17, 10.1145/3395351.3399421}. On the other hand, a controller deployed on a remote cloud could make the maintenance of the controller transparent to users and help with scalability and reliability concerns.

\subsection{\textcolor{black}{Security Analysis of the \sol Design}}

\noindent\textcolor{black}{We consider different potential attack scenarios that can take place in an IoT environment and we analyze the ability of \sol to mitigate these attack scenarios.}

\noindent\textcolor{black}{\textbf{Distributed Denial-of-Service (DDoS) attacks:} Due to vulnerabilities (such as weak authentication or encryption), IoT devices can be compromised by an attacker and used as a part of a botnet (\eg Mirai \textcolor{black}{~\cite{antonakakis2017understanding}}) to carry out DDoS attacks. In such scenarios, the \sol controller monitors traffic from and to the IoT devices. If unknown traffic patterns are detected, which do not match legitimate functions and communication among IoT devices, the controller will identify these patterns as anomalies and isolate the IoT devices causing them. As a result, \sol will be able to prevent the orchestration of DDoS attacks.}

\noindent\textcolor{black}{\textbf{Spoofing attacks:} If an IoT device gets compromised, {  an} attacker can spoof its IP address and/or its MAC address. As a result, the attacker can send malicious information to another IoT device within the IoT environment to manipulate its state or to a network outside the IoT environment. Since the \sol controller has a global view of the IoT environment (knowledge of MAC addresses and IP addresses of all devices), it can detect spoofed packets originating within the IoT environment and discard them.}

\noindent\textcolor{black}{\textbf{Impersonation attacks:} In this case, an attacker communicates with other devices as a legitimate device within the IoT environment and sends instructions to change the states of other devices. Because of the interaction trees created by \sol, impersonating devices will be identified and eventually isolated from the network.}

\noindent\textcolor{black}{\textbf{Passive attacks:} Passive attacks (such as traffic analysis, sniffing, eavesdropping) can be used to gather valuable information about activities (events) within an IoT environment without altering packets \cite{10.1145/3395351.3399421, DBLP:journals/corr/ApthorpeRF17,apthorpe2017closing}. \sol can defend against this type of attack by incorporating noise into the communication between the controller and IoT devices. The noise patterns will be agreed upon in advance between the controller and IoT devices. Since the controller will be aware of the noise patterns, it will be able to separate them from legitimate traffic.}


\subsection{Limitations of the \sol Design}

The \sol design has the following limitations, which we plan to address in our future work.


\noindent\textbf{Single point of failure:} The \sol controller introduces a single point of failure into an IoT environment. To mitigate this concern, two approaches can be used. The first approach involves maintaining multiple local IoT controllers in an IoT environment, since any IoT device with adequate resources can act as a controller. In this case, mechanisms to synchronize the state of local controllers will be needed. The second appraoch involves running a remote controller on a cloud. In this case, it would be the responsibility of the cloud provider to ensure that controller failure instances are effectively mitigated.



\noindent\textbf{Anomalies out of \solnospace’s scope:} \sol is designed to detect and recover from anomalies which are identified by monitoring network traffic and IoT device interactions. However, there are additional causes of anomalies, such as hardware or power failures of IoT devices, which \sol does not currently consider. {\revone For example, the design of \sol can be extended to detect anomalies of IoT devices using features of the physical layer. To achieve that, the controller will need to be equipped with additional hardware (\eg software-defined radio, spectrum analyzer) to collect physical layer data and our models will need to be trained accordingly in order to identify anomalies based on physical layer data.}






\noindent\textbf{Scalability of the \sol design:} 
In an IoT environment with hundreds {  of} IoT devices, such as a large building, \textcolor{black}{a large number of events and interactions may take place.} In such cases, the main challenge for \sol~{  would be to identify} devices with adequate resources, which can act as {  controllers}. \textcolor{black}{A solution could be to maintain multiple controllers and assign groups of IoT devices to different controllers}, so that the load is distributed among the available controllers.

\section{Conclusion and Future Work}
\label{sec:concl}

In this paper, we presented \sol, a framework that detects and mitigates the impact of traffic and interaction \noindent{anomalies among IoT devices.} Our evaluation results demonstrate that \sol is a lightweight framework that can identify IoT device anomalies in less than 2.12 ms and with up to 98\% of accuracy. In our future work, we plan to extend the \sol design in order to: (i) mitigate single point of failure concerns; (ii) identify additional categories of anomalies; (iii) support large-scale IoT environments with hundreds of IoT devices; and (iv) conduct additional experiments to further investigate the trade-offs of the \sol design.



\section*{Acknowledgements}

This work is partially supported by the National Science Foundation (awards CNS-2104700, CNS-2306685, CNS-2016714, and CBET-2124918) and ACM SIGMOBILE.

\bibliographystyle{IEEEtran}
\bibliography{sections/refs.bib}
\end{document}